\documentclass{article} 

\usepackage{graphicx,makeidx}
\usepackage{psfrag}
\usepackage{hyperref}

\usepackage{amsmath,bbm}

\newcommand{\bs}{\boldsymbol}
\newcommand{\R}{\mathbbm{R}}
\newcommand{\C}{\mathbbm{C}}

\newcommand{\dint}[1]{\int\frac{\rmd^d#1}{(2\pi)^d}} 

\newcommand{\tr}{\mathrm{tr}}
\newcommand{\rmi}{\mathrm{i}}
\newcommand{\rme}{\mathrm{e}}
\newcommand{\rmd}{\mathrm{d}}

\newcommand{\tauel}{\tau_\text{el}}
\newcommand{\gamel}{\gamma_\text{el}}
\newcommand{\tauphi}{\tau_\phi}
\newcommand{\elel}{\ell_\text{el}}
\newcommand{\Vsf}{V_\text{sf}}

\newcommand{\tausf}{\tau_\text{sf}}
\newcommand{\eltr}{\ell_\text{tr}}

\newcommand{\up}{{\uparrow}}
\newcommand{\down}{{\downarrow}}

\newcommand{\eps}{\varepsilon}

\newcommand{\ba}{{\boldsymbol{a}}}
\newcommand{\bb}{{\boldsymbol{b}}}
\newcommand{\bk}{{\boldsymbol{k}}}
\newcommand{\bn}{\boldsymbol{n}}
\newcommand{\bp}{{\boldsymbol{p}}}
\newcommand{\bq}{{\boldsymbol{q}}}
\newcommand{\br}{\boldsymbol{r}}
\newcommand{\bx}{\boldsymbol{x}}

\newcommand{\by}{\boldsymbol{y}}
\newcommand{\bB}{\boldsymbol{B}}
\newcommand{\bJ}{\boldsymbol{J}}
\newcommand{\bS}{\boldsymbol{S}}
\newcommand{\bT}{\boldsymbol{T}}

\newcommand{\ket}[1]{|#1\rangle}
\newcommand{\bra}[1]{\langle#1|}
\newcommand{\mv}[1]{\overline{#1}}
\newcommand{\expect}[1]{\left\langle#1\right\rangle}

\newcommand{\cg}[2]{\bra{#1}#2\rangle} 
\newcommand{\eq}[1]{\begin{equation} #1 \end{equation}}
\newcommand{\eqlab}[2]{\begin{equation} #1
        \label{#2.eq}\end{equation}}
\newcommand{\refeq}[1]{\textup{(\ref{#1.eq})}}

\newcommand{\tkq}{T^{K}_Q}
\newcommand{\tkqpr}{T^{K'}_{Q'}}

\newcommand{\calC}{\mathcal{C}}
\newcommand{\calV}{\mathcal{V}}
\newcommand{\calT}{\mathcal{T}}
\newcommand{\calD}{\mathcal{D}}

\newcommand{\calH}{\mathcal{H}}

\newcommand{\calL}{\mathcal{L}}
\newcommand{\calbarL}{\overline{\mathcal{L}}}

\newcounter{exo}
\newenvironment{cmexercise}[1]
{\stepcounter{exo} 
\small
\medskip \hrule\medskip
\noindent {\sffamily\textbf{Exercise \arabic{exo} --  #1}}} 
{\medskip \hrule \medskip}

\title{Diffusive spin transport
\protect\footnote{Notes based on a lecture given at the 
``International School on Quantum Information'', 
Max-Planck-Institut f\"ur Physik komplexer Systeme, Dresden, 2005} }
\author{Cord A. M\"uller}
\date{\small Physikalisches Institut, Universit\"at
Bayreuth, D-95440 Bayreuth, Germany}

\begin{document}

\maketitle

\begin{abstract} 

Information to be stored and transported requires physical carriers. The
quantum bit of information (qubit) can for instance be realised 
as the spin $\frac{1}{2}$ degree of freedom of a massive particle like
an electron or as the spin $1$ polarisation of a massless photon. In
this lecture, I first use irreducible representations of the rotation group
to characterise the spin dynamics in a least redundant
manner. Specifically, I describe the decoherence dynamics of an
arbitrary spin $S$  coupled to a randomly fluctuating magnetic field in the
Liouville space formalism. Secondly, I discuss the diffusive
dynamics of the particle's position in space due to the presence of
randomly placed impurities. Combining these two dynamics yields a
coherent, unified picture of diffusive spin transport, as applicable to 
mesoscopic electronic devices or photons propagating in cold atomic
clouds.  

\end{abstract}

\tableofcontents 

\newpage 

\section{Introduction}
\label{sec:outline}

Classical information processing uses charge encoding where 
the bit values ``0'' and ``1'' are represented by  a supplementary charge 
present or absent in a register. In the quantum limit, one eventually
is led to consider a single elementary charge (electron or hole) in 
a quantum dot  and hopes to realise quantum superpositions of
``qubit'' states
$\ket{0}$ and $\ket{1}$ and
entanglement between distinct qubits. But since they interact via 
long-range Coulomb
forces, charge states suffer strongly from decoherence. Another 
discrete degree of freedom is spin. Spin
$\frac{1}{2}$ states, typically noted $\ket{{\pm}1}$ or
$\ket{{\uparrow},{\downarrow}}$, are \textit{the} natural realisation of
a qubit. But spin, and more generally information as such, 
needs a physical carrier. Candidates here are electrons or holes 
(massive particle of spin $s=\frac{1}{2}$) and photons (massless, spin
$s=1$). A new promising field therefore is ``spint(r)onics'': 
spin-based information transport and processing with
elec\emph{trons} and pho\emph{tons}. 

\begin{figure}[b]
\begin{center}
\psfrag{0}{0}
\psfrag{L}{$L$}
\psfrag{p+0}{$p_\up(0)=1$}
\psfrag{p+L}{$p_\up(L)=?$}
\includegraphics[width=0.5\textwidth]{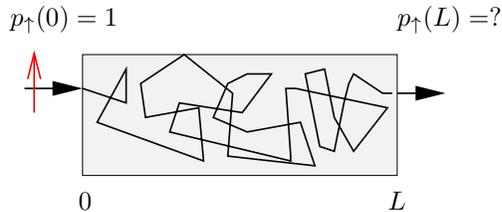}
\caption{\itshape 
\small Model setting of diffusive spin transport: spin-polarised particles are
injected from the left with $p_\up(0) =1$ into a disordered sample and
move diffusively towards
the right, where a spin-sensitive detection reads out the
spin-polarisation $p_\up(L)$. 
}
\label{spindiff.fig}
\end{center}   
\end{figure}

For these lectures,  I chose to discuss the following model
setting of diffusive spin transport (Fig.\ \ref{spindiff.fig}): spin-polarised particles are
injected from the left with probability $p_\up(0) =1$ into a disordered sample and
move diffusively towards 
the right, where a spin-sensitive detection reads out the final 
spin polarisation $p_\up(L)$ that we should calculate. 

The leitmotiv of this lecture was: Spin is a geometrical quantity, and
one should be able to use irreducible
representations of the rotation group in order to take advantage of
symmetries. Since the participants of the school were not
required to have been educated in group theory, I decided to give a rather
complete introduction into representations of the rotation group with
the hope that the sometimes intimidating jargon of group-theoretical
arguments may become more familiar to everybody. Attending myself
other lectures, I adopted a master-equation approach parallelling Klaus
Hornberger's lecture \cite{klh06} to which the present notes refer
occasionally. During the lecture itself, only the first part on spin
relaxation was delivered as presented in the following; in the second
part on diffusion, I relied on arguments taken from diagrammatic
perturbation theory that seemed to bewilder the audience more than
anything else. The present notes  remedy to this
dissymmetry and treat also the diffusive part as momentum relaxation with a
master equation. This parallel allows to combine both dynamics rather
economically into a single, coherent picture of diffusive spin
dynamics; it is my hope that this conceptual unity is appreciated by my readers. 

These notes finish with a description of quantum corrections to
diffusive spin transport including
spin-flip effects. Other subjects covered during the school's lecture
were of a more anecdotic type and, although hopefully enjoyed by the audience, 
did not seem to fit into the present format; readers interested in 
spintronics properly speaking are referred to the recent review by $\check{\mathrm Z}$uti\'c,
Fabian, and Das Sarma \cite{Zutic04}. I have to admit that I made no attempt
to cover systematically the vast literature on the subject of irreversible
spin dynamics and quantum transport, which would have
been a hopeless task in any case; my
apologies to many colleagues whose excellent contributions may not be
duly cited in the following.

\section{Spin relaxation}
\label{spinrelaxation.sec}

``Spin'' is internal angular momentum  \cite{CH,BL81}. This was
recognised by Uhlenbeck and Goudsmit 
\cite{UG25} 
following Pauli 
\cite{Pauli25} who postulated the existence of a fourth quantum number
to explain fine-structure features of atomic spectra. The clearest
experimental manifestation of quantised spin is arguably 
the Stern-Gerlach experiment 
\cite{GS22} where silver atoms are deviated by an inhomogeneous
magnetic field into two distinct spots on a detector screen.  

Dirac discovered that bispinors (vectors of four components, a spin
$\frac{1}{2}$ 
spinor and an anti-spinor) appear naturally when one looks for a
Schr\"odinger-type wave equation in the relativistic framework 
of the four-dimensional Minkowski space.   
Wigner  
\cite{Gross95} showed that spin
is one of the fundamental quantum numbers that permits to identify an
elementary particle in the first place: 
it characterises the particle's properties under 
rotations in its proper rest frame. Therefore, to understand spin
is to understand rotations. 

\subsection{Spin -- a primer on rotations}

\subsubsection{Rotation group}
\label{rotationgroup.sec}

Physical objects are described by coordinates $\bx=(x_1,x_2,x_3) \in
\R^3$ with
respect to a reference frame in configuration space. 
Rotations (called ``active'' when the
object is turned and ``passive'' when the reference frame is turned)
are represented by $3\times 3$ matrices: 
$\bx' = R \bx$. Proper rotations conserve the
Euclidean scalar product $\bx\cdot\by=\sum_ix_iy_i$ and the
orientation of the frame. The rotation matrices are therefore members of SO(3),
the set of orthogonal matrices $RR^\text{t} =
R^\text{t}R=\mathbbm{1}_3$ of unit determinant $\det R= +1$. 
With the usual matrix
multiplication as an internal composition law, theses matrices form a
\emph{group}, satisfying the group axioms: 
\begin{enumerate}
\item 
Internal composition: $\forall R_{1,2} \in \text{SO(3)}: 
R_{21}=R_2R_1 \in \text{SO(3)}$;
\item 
Existence of the identity: $\exists E: R E =E R =R \ \forall
  R\in\text{SO(3)}$ 
with $E=\mathbbm{1}_3$; 
\item 
Existence of the inverse: $\forall R \exists R^{-1}: R R^{-1}=
R^{-1}R=E$. 
\end{enumerate}
The group is \emph{non-Abelian} because the matrices do not commute:
$R_2R_1\neq R_1R_2$. An exception are rotations of the plane around
one and the same axis, forming the Abelian group SO(2).

\begin{figure}[b!]
\begin{center}
\includegraphics[width=0.3\textwidth]{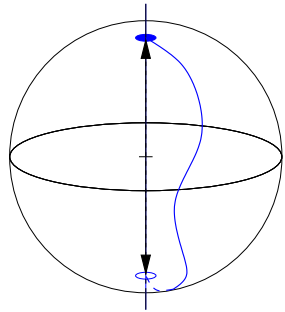}
\caption{\itshape\small The parameter space of the rotation group SO(3), a filled sphere of radius
$\pi$. Two identical rotations 
[eq. \refeq{identrot}] can be connected by a closed curve that is not
reducible to a single point.}
\label{so3sphere.fig}
\end{center}   
\end{figure}

A possible parametrisation of a rotation is
the polar description $(\hat\bn,\theta) =: \bs\theta$ 
with $\hat\bn$ the unit vector along
the rotation axis and $\theta\in[0,\pi]$ the rotation angle. The
following two rotations of configuration space are identical: 
\eqlab{
R(\hat\bn,\theta=\pi) = R(-\hat\bn,\theta=\pi). 
}{identrot}
These opposite
points must be identified such that there are closed parameter curves
that cannot be contracted into a single point, see figure
\ref{so3sphere.fig}. This means that SO(3) is a
doubly connected manifold. Instead of studying this
projective group, one may also turn to its \emph{universal covering
group} SU(2), the group of all unitary $2\times 2$ matrices over $\C$
with unit determinant. SU(2) is simply connected, and there is a
\emph{homomorphism} (a mapping preserving the group structure) linking
every
element $U\in$ SU(2) to the rotation $R\in$ SO(3): the rotation
$\bx'=R(\bs\theta) \bx$ is described by   
\eqlab{
\bx'\cdot\bs\sigma = U(\bs\theta)\bx \cdot\bs\sigma
U(\bs\theta)^\dagger.
}{Uaction.rot} 
Here, $\bs\sigma=(\sigma_1,\sigma_2,\sigma_3)$ is a vector whose
entries $\sigma_i$ are the Pauli
matrices such that 
\eq{
\bx\cdot\bs\sigma = \begin{pmatrix} x_3 		& x_1-\rmi x_2 \\
				x_1 + \rmi x_2 	& -x_3 
		 \end{pmatrix}
 }
and the  unitary rotation matrix acting from the left and from the
right is given by 
\eqlab{
U(\bs\theta) =  \mathbbm{1}_2 \cos\frac{\theta}{2} - \rmi
\hat\bn\cdot\bs\sigma \sin\frac{\theta}{2}= \exp (-\rmi \frac{\bs\theta\cdot\bs\sigma}{2}). 
}{su2param} 
Note that this mapping is two-to-one because $U$ and $-U$ yield the same
rotation, and that it takes a rotation by an angle of $4\pi$ to recover the identity
transformation:  
$U(\theta=2\pi) = - \mathbbm{1}_2$, but $U(\theta=4\pi) = +
\mathbbm{1}_2$. This is not a mysterious quantum property as is
sometimes stated, but reflects the double-connectedness of
SO(3)-rotations in our everyday reference frame.%
\footnote{Cartan developed 
half-integer spin representations as early as 1913 within the theory
of projective groups \cite{Cartan13}.}
Dirac's construction of a solid body connected
by strings to a reference frame is supposed to convey an
``experimental'' idea of this property \cite{BL81}.

\subsubsection{Representations}

A group $G$ can act in many different disguises that share the same
abstract group structure, as defined by the multiplication law or
group table. These different appearances are called (linear) 
\emph{representations}. Mathematically, they are mappings $\calD_i: G \to GL(V_i)$  from the group $G$ to the
general (linear) group of  regular transformations $D: V_i \to V_i$ 
 of a vector space $V_i$ into itself.%
\footnote{$\calD$ stands for the German word
 ``Darstellung''; in anglo-saxon literature, often the symbol $\Gamma$
is used.} 
Importantly, this mapping must be a 
\emph{homomorphism} which means that the representation has
 the same group structure as $G$. Notably, for all elements
 $g_1,g_2\in G$ (with the product $g_2g_1\in G$) the representing
transformations verify $D(g_2g_1)=
 D(g_2)D(g_1)$.

$\dim V_i$ is called the dimension of the
 representation $\calD_i$. Finite-dimensional linear representations
 are given in terms of quadratic matrices  of size $\dim V_i$. 
A representation is called \emph{reducible} if there is a basis of
$V=V_1\oplus V_2$ such that all transformations $D \in \calD$ are
written 
\eq{
D = \begin{pmatrix}
 D_1 & \ast \\
 0 & D_2 
\end{pmatrix}. 
} 
In other words: the transformations $D_1: V_1 \to V_1$ define already a
representation $\calD_1$ of its own. If the matrix is block-diagonal
($\ast=0$), the representation is \emph{completely reducible}. If it
is not reducible, it is called \emph{irreducible}, meaning that one
has achieved to work in the smallest possible subspace.

The rotation group SU(2) is a manifold that 
depends on a continuous set of parameters 
$\bs\theta=(\theta_1,\theta_2,\theta_3)$ with respect to which it is 
infinitely differentiable. This structure is called a 
\emph{Lie  group}. The group multiplication law is completely 
determined by the commutation
relation of its generators: 
\eqlab{
[J_j,J_k]:= J_jJ_k - J_k J_j =  \rmi \hbar  \epsilon_{jkl} J_l. 
}{commutationsu2}
Here as in the following, the sum over repeated indices is
understood. 
These generators are said to form a (representation of) the
\emph{Lie algebra} $\mathsf{su}(2)$.  In the 
so-called natural representation of SU(2) by itself
[eq.~\refeq{su2param}], the generators are $\bJ = \hbar 
\frac{\bs\sigma}{2}$. Other representations will feature different
generators, but 
all group representations share the same commutation relation! 
Finite rotations are generated by exponentiation: 
$
U(\bs \theta) =\exp(-\rmi \bs \theta\cdot\bs J/\hbar )
$.
Topologically speaking, SU(2) is compact. According to a general theorem, all
representions of a compact Lie group are completely reducible to
finite-dimensional irreducible representations.

\subsubsection{Functional representation of SO(3)} 
\label{functrep.sec} 

As an example for an infinite-dimensional representation of the
rotation group SO(3) already useful in classical physics, consider the
transformation of functions $f$ describing the position of an object
on the unit sphere. Saying that the object is rotated to $\br'=R\br$
implies that the function is transformed by 
\eq{
(Df)(\br) = f(R^{-1}\br). 
} 
The action of $D$ in the (infinite-dimensional) functional space can
be written $D= \exp\{ -\rmi \bs\theta \cdot\bs T\}$ where $\bs T=
-\rmi \bs r\times\bs\nabla$ is a differential operator, as can be
verified by considering an infinitesimal rotation  around $\bs\theta = \theta
\hat{\bs{n}}$: to first order in $\theta$, we have $f(R^{-1}\br) =
f(\br - \bs\theta\times\br) = f(\br) -
(\bs\theta\times\br)\cdot\bs\nabla f(\br) =
[1-\bs\theta\cdot(\br\times\bs\nabla)]f(\br)$. 

The  finite-dimensional irreducible representations to which this
infinite-di\-men\-sional one can be reduced are obtained by decomposing
$f$ into surface 
harmonics $Y_{Lm}$; each subspace $L=0,1,\dots$ then
admits an irreducible representation of dimension $2L+1$. 
Surface harmonics are a concept arising already
with the multipole expansion
of charge distributions in classical electrodynamics. But to cite Hermann Weyl \cite{Weyl50}: ``This reveals the true significance of
surface harmonics; they are characterised by the fundamental symmetry
properties here developed, and the solution of the potential equation
in polar co-ordinates is merely an accidental approach to their
theory.''

\subsubsection{What is quantum about spin?}

Spin, and especially half-integer spin, is much less mysterious than 
sometimes suggested by standard textbook wisdom. Half-integer spin does
not require the kinematic framework of special relativity, it arises
already in Galilean relativity. Also, as mentioned at the end of \ref{rotationgroup.sec}, reference frames
for solid bodies already introduce half-integer spin. However, there are genuine quantum
features to be aware of: 

\begin{enumerate}
\item 
In classical mechanics, Noether's theorem assures that if the
Lagrangian function is invariant under infinitesimal rotations, then the orbital angular
momentum $\bs L$ is a conserved quantity. However, $\bs L$ a priori 
has nothing to do with the generator of rotations $\bs T$  introduced in \ref{functrep.sec}. 
In quantum mechanics, thanks to the appearance of $\hbar$, the
generators can be \emph{identified} with \emph{observables} $\bJ=\hbar\bs T$ with dimension of angular
momentum. If the Hamiltonian is invariant under all rotations $U(\bs\theta)$, 
  their generators itself are conserved
  quantities, and Noether's theorem takes a very simple form:  
\eq{
H'= U H U^\dagger \Leftrightarrow [H,U]=0 \Leftrightarrow
[H,\bJ]=0 \Leftrightarrow  \dot{\bJ}= 0.    
}

Separating the orbital part $\bs L$ of angular momentum from
the total angular momentum $\bs J = \bs L \otimes \mathbbm{1}_S + \mathbbm{1}_L \otimes
\bS =: \bs L +\bS$, one identifies the rest-frame angular momentum or
spin $\bS$ obeying the same fundamental commutation relations
\refeq{commutationsu2}. In the remainder, we will only have to
consider the spin part and let $\bs L=0$. 

\item The observables $\bs S$ generate irreducible
  representations $\calD^{(s)}$ of dimension $d_s=2s+1$ 
  with $s=0,\frac{1}{2},\dots$ and discrete
  magnetic quantum numbers $m=-s,-s+1, \dots,s$. 
Pure states are noted $\ket{s m}$. The \emph{Casimir operator} $S^2$
specifies the irreducible representation, $S^2\ket{s m}=
\hbar^2 s(s+1)\ket{sm}$, whereas the magnetic quantum number gives the
projection of the spin onto the quantisation axis (usually called the
$z$-axis): $S_z\ket{sm} = \hbar m\ket{sm}$.     

\item 
States can be classified
regarding their transformation properties, but here with
a more general importance than in classical mechanics 
due to the superposition principle. An atomic 
$s$-orbital, for instance, may be seen as a ``coherent superposition of all
possible Kepler orbits'' and is invariant under all rotations. 
In section \ref{statesmultipoles.sec}, we
will introduce the irreducible components of mixed quantum states,
also known as \emph{state multipoles}.

\end{enumerate}

\subsection{Master equation approach to spin relaxation}
\label{sec:spindyn}

As a simple model for spin dynamics, we shall study the Hamiltonian 
\eqlab{
H=- \mu \bS \cdot \bB\ .
}{HSB} 
It describes the coupling of the magnetic
moment $\bs \mu =\mu \bS $ to a magnetic field $\bB$. For electrons, 
$\mu= -g \mu_\text{B}$ in terms of the Bohr magneton
$\mu_\text{B}=|e|\hbar/(2m_e c)$ and the gyromagnetic ratio
$g=2.003...$ in vacuum. 
The spin operator has been chosen dimensionless such
that its action 
in the irreducible representation $\calD^{(s)}$  will
be $S_z\ket{sm}= m\ket{sm}$ and $S^2\ket{sm} = s(s+1)\ket{sm}$ for 
the remainder of the lecture. 
The density matrix or statistical operator $\rho$ of a spin $S$ is a positive
linear operator of trace unity on the state 
Hilbert space $\calH_s=\C^{d_s}$ of dimension $d_s:=\dim\calH_s= 2s+1$
that determines the expectation values of abitrary observables
$O$ as $\expect{O} = \tr\{\rho O\}$.

\subsubsection{Unitary spin dynamics}
\label{unitary.sec}

According to one of the fundamental axioms
of quantum theory, any closed quantum system evolves unitarily
according to the 
Liouville-von Neumann  equation 
\eqlab{
\rmi \hbar \partial _t \rho =[H,\rho].
}{unitary}  
This equation of motion is formally solved as $\rho(t_2) = U(t_2,t_1) \rho(t_1) U(t_2,t_1)^\dagger$
by applying the time evolution operator for a time-dependent
Hamiltonian, 
\eq{
U(t_2,t_1) = \mathrm{T} \exp\left\{ - \frac{\rmi}{\hbar}
\int_{t_1}^{t_2} H(t')\rmd t'\right\} 
} 
where $\mathrm{T}[H(t_1)H(t_2)\dots H(t_n)] = H(t_i) H(t_j) \dots
H(t_k)$ for $t_i > t_j >\dots  > t_k$ is Dyson's time-ordering operation.

\subsubsection{Non-unitary spin dynamics: a classical model derivation}
\label{nonunitspin.sec}

Phenomena like ``relaxation, damping, dephasing, decoherence,...'' have in common 
irreversible dynamics with an ``arrow of time'' \cite{Zeh} due to the
irrevocable loss of energy and/or information into inobservable
degrees of freedom which are usually called ``bath'' or ``environment''. 

As an introductory model, we consider the Hamiltonian 
$H=-\mu \bs S\cdot \bB(t)$ with a randomly fluctuating
magnetic field $ \bB(t)$. Predictions about the spin will involve an
average over the field fluctuations which we describe as a 
\emph{classical} stochastic process \cite{vKampen}. 
This approach is typical for the physics of nuclear magnetic
resonance; a regular driving magnetic field can of course be included
in the treatment, but here we concentrate on the effect of random
fluctuations.

To obtain the effective dynamics, we develop the time-propagated 
density matrix
$\rho(t+\Delta t)=U(t+\Delta t,t)\rho(t)U(t+\Delta t,t)^\dagger$ to second 
order in the interaction
Hamiltonian, 
\eqlab{
\begin{aligned}
\rho(t+\Delta t) & =  \rho(t) - \frac{\rmi}{\hbar} \int\limits_t^{t+\Delta t}
[H(t_1),\rho(t)]  \rmd t_1 + \frac{1}{\hbar^2} \int \!\!\!\!\!\!\int\limits_t^{t+\Delta t} 
    H(t_1)\rho(t)H(t_2) \rmd t_1\rmd t_2 \\
 &  \qquad \quad  -\frac{1}{2\hbar^2} \int \!\!\!\!\!\!\int\limits_t^{t+\Delta t} \text{T} (H(t_1)H(t_2)\rho(t) + \rho(t)  H(t_1)H(t_2)) \rmd t_1\rmd
 t_2, 
\end{aligned}
}{rhodeltat}
and perform an average over all possible realisations of the
fluctuating magnetic field $\bB(t)$. As a stochastic process
\cite{vKampen}, it is completely specified by its correlation
functions 
\eqlab{
C_{i_1i_2 \dots i_n}(t_1,t_2, \dots,t_n):= 
\mv{B_{i_1}(t_1)B_{i_2}(t_2)  \dots 
B_{i_n}(t_n)}, 
}{Bcorr}
the overline indicating an ensemble average over 
the field distribution. 
This distribution is 
taken to be centered Gaussian 
such that all correlation functions factorise into
products of pair correlations. Therefore, 
only the first two moments need to be
specified: $\mv{B_i(t)} = 0 $  (zero mean) and 
\eq{
\mv{B_i(t_1)B_j(t_2)} =: B^2 c_{ij}(t_1,t_2).
} 
We assume a stationary process that depends only on
the time difference $t_1-t_2$ and has a very short internal correlation time
$\tau_c$ such that $c_{ij}(t)=c_{ij} \tau_c \delta(t)$. This last
assumption of ``white noise'' (the
power spectrum $\tilde c_{ij}(\omega) = cst.$ contains all frequencies
with equal weight) is valid if the noise correlation time
$\tau_c$ is much shorter than the relevant time scale of the spin
dynamics that is still to be determined.  
Lastly, we assume that the fluctuations are
\emph{isotropic}, $c_{ij}=\frac{1}{3}\delta_{ij}$. 

Now we average the time-propagated density matrix \refeq{rhodeltat}
over the field
fluctuations and use the \emph{Born assumption} that there are no
further correlations between the
fields appearing explicitly  and the average density matrix. Then, the
ensemble average applies to the fields only. 
The term linear in $H$ disappears because $\mv{B_i}=0$. 
In the second order terms, one of the time integrations is
contracted by the $\delta(t_1-t_2)$-distribution of the correlation function; the
remaining integrand is time independent such that the integral gives
just a factor $\Delta t$. The average time-evolved density matrix 
then becomes 
\eqlab{
\rho(t+\Delta t) = (1-\gamma_s\Delta t) \rho(t) + \gamma_s\Delta
t \sum_i \hat S_i \rho(t) \hat S_i + O((\gamma_s\Delta t)^2)
 }{rhotplus} 
where $\hat S_i := S_i/\sqrt{s(s+1)}$ is the ``normalised spin
operator'' with $\sum_i\hat S_i^2 = \mathbbm{1}$. 
The spin relaxation rate $\gamma_s:= s(s+1) \omega_0^2 \tau_c$ is given
in terms of the 
 squared effective Larmor frequency $\omega_0^2 = \mu ^2 B^2/(3\hbar^2)$.
The relevant time scale of evolution turns out to be $\tau_s:=1/\gamma_s$. The effective
time-evolution \refeq{rhotplus} then is valid for a small time step $\Delta t\ll
\tau_s$ such that indeed $\gamma_s\Delta t\ll 1$.

\begin{cmexercise}{Non-unitary spin dynamics: quantum derivation}
\label{nonunitspin_quantum.exo}

Consider the time-independent model Hamiltonian 
$H=- \hbar J \bs S\cdot \bs \tau $
where our spin $\bS$ is coupled to a freely orientable magnetic
impurity, here modeled as a spin $\frac 1 2$ with Pauli
matrices $\bs \tau$. The effective
spin dynamics of $\bS$ 
is described by its reduced density matrix $\rho(t)
=\tr_\tau\{\rho_{S\tau}(t)\}$ obtained by tracing out the
uncontrolled impurity spin.  
Develop the
time-evolved complete density matrix $\rho_{S\tau}(t)$ to
second order in $J$ as in the previous section \ref{nonunitspin.sec} 
and take the trace over the impurity with initial statistical mixture $\rho_\tau(0) =
\frac{1}{2} \mathbbm{1}_2$ (it is helpful to use the identities 
$\tr_\tau \{\tau_i\} =0$ and $\tr_\tau \{\tau_i\tau_j\} = 2
\delta_{ij}$). 
Show that one finds exactly the 
evolution \refeq{rhotplus} with a relaxation rate $\gamma_s = s(s+1) J^2 \Delta
t$ up to higher-order terms that become negligible in the formal limit $\Delta t\to 0$
together with $J\to \infty$ taken at constant $\gamma_s$.%
\footnote{But
attention: at finite coupling $J$, the dynamics of our spin shows 
\emph{recurrence} on a time scale given by the so-called Poincar\'e time
$t_\text{rec}\propto 1/J$. One could obtain a truly irreversible 
dynamics only by supposing that the single impurity spin is reset
rapidly enough in order to dispose of the
coherence. Alternatively, one may imagine the setting treated in section
\ref{spindiffusion.sec}: our spin is moving and encounters different
impurity spins such that in the thermodynamic limit the Poincar\'e
recurrence time goes to infinity and true irreversibility sets in.}

\end{cmexercise}

\subsubsection{The quantum channel and its operator sum representation}
\label{quantumchannel.sec}

In the language of quantum information, the time evolution
\refeq{rhotplus} up to order 
$\Delta t$ defines a
\emph{quantum channel} $\rho = \rho(t) \mapsto \rho' = \rho(t+\Delta t)$
and  is here given in the so-called  \emph{operator sum
representation}  \cite[8.2.3.]{NielsenChuang} 
\eqlab{ 
\rho' = \sum_{i=0}^3 W_i \rho W_i^\dagger
}{opsumrep} 
with 
\eq{
W_0:= \sqrt{\smash[b]{1-\gamma_s\Delta t}}\,\mathbbm{1}, \qquad 
W_i:=\sqrt{\smash[b]{\gamma_s\Delta t}}\,
\hat S_i,\quad i=1,2,3. 
} 
It is easy to verify that 
$\sum_{i=0}^3 W_i W_i^\dagger =\mathbbm{1}$
which guarantees the trace
conservation $\tr \rho'=\tr \rho$. 
Kraus has proved that a channel of this form assures that the final density matrix is again
completely positive. Therefore, it is also known as the \emph{Kraus
representation} and the $W_i$ are commonly referred to as 
\emph{Kraus operators}, cf.\ sec.\ 3.1 in \cite{klh06}.  
In contrast to the unitary evolution $\rho' = U\rho U^\dagger$ of \refeq{unitary}, the
appearance of several independent terms in the sum \refeq{opsumrep}
signals non-unitary dynamics. 

For a spin $\frac{1}{2}$ with $\hat S_i=\sigma_i/\sqrt{3}$, this
quantum channel is 
the  \emph{qubit depolarising channel} \cite[8.3.4.]{NielsenChuang}: 
\eq{
\rho' = (1-p_1) \rho + \frac{p_1}{3} \sum_i \sigma_i \rho \sigma_i ,\quad
p_1=\gamma_s\Delta t. 
}
With equal 
probability $p_1/3$, the qubit is affected by the action of one of
the Pauli matrices  $\sigma_i$, and with probability $1-p_1$, it remains
untouched. Since for spin $\frac{1}{2}$ one may write 
$\sum_i \sigma_i \rho \sigma_i = 2 \mathbbm{1}_2 - \rho$, the
channel also takes the suggestive form 
\eqlab{
\rho' = (1-p_2) \rho + \frac{p_2}{2} \mathbbm{1}_2. 
}{depolarchannel}
This means that with probability $p_2 = 4p_1/3$ (remember $p_1\ll 1$ such that also $p_2\ll
1$), the density matrix is taken to a complete statistical mixture and
remains identical with probability $(1-p_2)$. 

By convention, the ``depolarising channel'' for higher spin 
$S\ge 1$ (i.e. $d_s\times d_s$ density matrices with $d_s=2s+1 \ge 3)$ is
still defined via the relation \refeq{depolarchannel} with the
statistical mixture $\frac{1}{d_s} \mathbbm{1}_{d_s}$ on the
right-hand side.
This channel is also called ``SU($n$) channel'' with $n=d_s$ because 
the corresponding Kraus operators are the $n^2-1$  
generators of the Lie algebra $\mathsf{su}(n)$ 
\cite{Ritter05}.  
Note that our physical model of an arbitrary spin $\bS$ coupled to a
fluctuating magnetic field does \emph{not} lead to this specific Lie
algebra channel: obviously, the operator-sum representation
\refeq{opsumrep} contains only the 3 generators of $\mathsf{su}(2)$,
albeit in a representation of dimension $d_s= 2s+1$. We will see in
the following that this makes the spin dynamics richer and its
description more involved. Group-theoretical methods will be
introduced that are  well 
adapted to cope with this complexity.

\subsubsection{The Liouvillian}
\label{Superoperators}

The linear operators on the state Hilbert space $\calH_s$ are themselves elements of a linear
vector space (we can add operators and multiply them by complex
numbers). 
This vector space is called \emph{Liouville space} $L(\calH_s)$ and
is spanned, for example, by the basis of \emph{dyadics} induced by
 basis vectors $\ket{n}$ of $\calH_s$: 
\eq{
\ket{m}\bra{n}=: |mn),\quad n,m=1,\dots,d_s. 
}
The Liouville-von Neumann equation of motion 
$\rmi\partial _t \rho =\calL\rho$ for a closed quantum system 
defines the \emph{Liouvillian} 
\eqlab{
\calL = \frac{1}{\hbar} [H,\cdot]  
}
{defLiouvill} 
whose matrix elements in the
dyadic basis are 
\begin{equation}
(mn|\calL|m'n') = \calL_{mn,m'n'} = \frac{1}{\hbar} \left(H_{mm'}\delta_{nn'} -
H_{n'n}\delta_{mm'}\right). 
\end{equation}    

\begin{cmexercise}{Liouvillian eigenvalues}

Show that the eigenvalues of the Liouvillian
in the basis $\{|mn) \}$ induced by the energy basis
$H\ket{m}=\eps_m\ket{m}$ are the possible transition
frequencies $\omega_{mn} = (\eps_{m}-\eps_n)/\hbar$. These are
experimentally accessible quantities, in contrast to the absolute
energy eigenvalues of the Hamiltonian $H$.  

\end{cmexercise}

As an operator between operators, the Liouvillian $\calL: 
L(\calH_s)\to L(\calH_s)$ is called a \emph{superoperator}
\cite{FS,Zwanzig64,Gabriel,Preskill}.   
The superoperator formalism is a convenient starting point for projection
operator techniques in statistical dynamics \cite{Zwanzig64}, effective
dynamics of open quantum systems \cite{GZ04} and time-dependent
perturbation theory \cite{Mukamel04}.

\subsubsection{The Lindbladian}
\label{lindbladian.sec} 

The effective dynamics of our spin coupled to a randomly fluctuating
field can also be formulated in terms of a superoperator. 
Taking the formal limit $\lim_{\Delta t \to 0}\frac{\rho(t+\Delta t) -\rho(t)}{\Delta t}=: \partial_t \rho(t)$
in \refeq{rhotplus} leads to the \emph{master equation}
\eqlab{
\partial_t \rho(t) = \overline{\calL}\rho(t) , 
}{Lbarrho}
a linear
equation of motion for the effective density operator,  whose effective
generator of time evolution is 
the \emph{Lindbladian} 
\cite{Preskill}
\eqlab{
\overline{\calL}\rho(t) = -\frac{\gamma_s}{2}\sum_i[\hat S_i,[\hat S_i,\rho(t)]]
= \frac{\gamma_s}{2}\sum_i \left ( 2 \hat S_i \rho(t)\hat S_i- \hat S_i
\hat S_i \rho(t)- \rho(t)\hat S_i \hat S_i\right).  
}{Lspinrelax}
This is a pure relaxation superoperator in the Lindblad
form, 
which assures the complete positivity of
the time-evolved density matrix. 
The symbolic limit $\Delta t\to 0$  really means $\Delta
t \gamma_s \ll 1$ but is still assumed to be  ``coarse-grained'' 
compared to the field correlations, $\Delta t \gg \tau_c$. 
In this limit, the master
equation \refeq{Lbarrho} is a linear equation for the density matrix
$\rho(t)$ local in time $t$ and thus describes Markovian dynamics without any memory
effects.  

\begin{cmexercise}{Lindbladian matrix elements} 

Show that the superoperator matrix elements of
the Lindbladian \refeq{Lspinrelax} are given by 
\eqlab{
\overline{\calL}_{mn,m'n'} =\gamma_s\left(  \bs{\hat{S}}_{mm'}\cdot \bs{\hat{S}}_{n'n} -
\delta_{mm'} \delta_{nn'}\right) 
}{calL.def}
and verify the trace-preserving property
$\sum_m\overline{\calL}_{mm,m'n'}=0$ 
from this expression. 
\end{cmexercise}

\subsection{Irreducible scalar spin relaxation rates}
\label{sec: irreducibleSpops}

Formally, the solution of the master equation \refeq{Lbarrho} is very
simple: 
\eqlab{
\rho(t) = \exp[\calbarL t]\rho(0).
}{solutionmastereq} 
The dynamics induced by the Lindbladian is called a ``quantum dynamical
semigroup'' (Sec. 3.1 in \cite{klh06}). Indeed, the time evolution superoperator satisfies
$\exp[\calbarL (t_2+t_1)] = \exp[\calbarL t_2]\exp[\calbarL t_1]$ and
$\exp[\calbarL 0] = \mathbbm{1}$ which indicates a group structure. 
``Semigroup'' means that the inverse to each element does not need to exist,
and indeed here it does not since the non-unitary dynamics
obtained by tracing out the environment has induced an arrow of
time. The Kraus representation \refeq{opsumrep}  and the
Lindblad form \refeq{Lspinrelax} guarantee the complete positivity of
the final density matrix if the initial one is completely positive,
but the inverse is not true: the Lindbladian is not invertible
\cite[3.4.1.]{Preskill}. 
We will indeed see in section \ref{eigenvalues.sec} below that $\calbarL$  has 
one vanishing eigenvalue $\lambda_0=0$. 

In Liouville space, the master equation is a matrix equation
$\partial_t|\rho) = \calbarL |\rho)$   
of dimension $d_s^2\times d_s^2$. As always when dealing with matrix
equations,  we have to diagonalise the
Lindbladian in Liouville space in order to be able to use the
formal solution \refeq{solutionmastereq}. 
For spin $\frac{1}{2}$ and $d_s=2$, diagonalizing a $4\times 4$ matix
is elementary and can be done by hand, but already for spin 1 with $d_s^2=9$ this becomes
cumbersome.  For higher spin, one definitely needs to resort to an
efficient strategy to find the eigenstructure of the Lindbladian.  
In the following, we show how to take maximal advantage of rotational
symmetries by using group-theoretical methods that lead to a very
simple and physically transparent description for the effective spin
dynamics. 
 
\subsubsection{Scalar relaxation process: what results we should expect}

The Lindbladian was obtained by an isotropic average and is thus a
\emph{scalar} object, i.e., invariant under rotations. A 
rather high-brow  way of expressing this simple property is to say ``it transforms under
the trivial representation $\calD^{(0)}$''. We may anticipate
that the statistical operator can be decomposed   into parts 
that transform under the irreducible representations
$\calD^{(K)}$ of the rotation group, 
$\rho = \sum_K \rho^{(K)}$. The Lindbladian as a scalar
object can only connect subspaces of equal rank $K$. 
Furthermore, inside each subspace it cannot distinguish between
different orientations. Thus, 
in an adapted basis of Liouville space,  
it can be written as a purely diagonal matrix 
\eq{
\calbarL = \begin{pmatrix} 
  \lambda_0 &  0 & \dots & 0\\
  0         & \lambda_1  & \dots &0    \\
 \vdots & \vdots & \ddots  & \vdots \\
 0 &  0 & \dots  & \lambda_{d_s^2}  
   \end{pmatrix} 
}
How many \emph{different} eigenvalues may we expect? The total number is the
dimension of the Liouville space, $d_s^2$. Each subspace of rank $K$ will have
dimension $d_K = 2K+1$. Therefore, we will have to find only $2s+1$
different eigenvalues $\lambda_K$, each of which has degeneracy 
$2K+1$: 
\begin{center}
\renewcommand{\arraystretch}{1.5}
\begin{tabular}{r| c c c l l} 
 particle  & $s$   &  $d_s$   &  $d_s^2$  & eigenvalues   &  degeneracy
 \\
\hline
 electron &  $\frac{1}{2}$  & 2 & 4 &  $ \lambda_0, \lambda_1$ & 1, 3 \\
 photon &  $1$  & 3 & 9 &  $\lambda_0, \lambda_1, \lambda_2 $ & 1, 3, 5 
\end{tabular}
\end{center}
Can some of the eigenvalues be identical which would imply an even larger
degeneracy? A trivial example for this would be any operator
proportional to the identity. But we will see below that this is not
the case for the 
Lindbladian \refeq{Lspinrelax}: the eigenvalues pertaining to
different subspaces $K\neq K'$ are indeed different, $\lambda_K\neq
\lambda_{K'}$. We are therefore sure to have reduced the
problem to the simplest possible formulation.

\subsubsection{Irreducible tensor operators}

Before we can define irreducible superoperators, we first had better
understand the simpler concept of ordinary irreducible operators. 
An irreducible tensor operator by definition is a set of $2K+1$ components $\tkq$, $Q=-K, -K+1, \dots, K$, that
transform under irreducible representations of rank $K$ (i.e., whose
transformation does not mix different $K$):  
\eq{
(\tkq)' = U \tkq U^\dagger = \sum_{Q=-K}^K D^{(K)}_{QQ'}T^{K}_{Q'} . 
} 
Equivalently, one specifies the infinitesimal rotation properties by
requiring 
\begin{eqnarray}
\, [J_\pm,\tkq] & = & \hbar \sqrt{K(K+1)-Q(Q\pm1)} T^{(K)}_{Q\pm 1},  
\label{ITOcom1.eq}\\
\, [J_0,\tkq]  & = &\hbar Q \,\tkq .
\label{ITOcom2.eq} 
\end{eqnarray}
Here, the angular momentum raising and lowering operators are $J_\pm =
J_x\pm i J_y$. 
The simplest examples for irreducible tensor operators we need to know
for the following are 
\begin{itemize}
\item[-] $K=0$ or scalar operator $T^0_0$, a single operator that commutes with all
components of the total angular momentum $\bs J$. For instance $S^2$,
the Casimir operator indexing the irreducible representations
$\calD^{(s)}$. 

\item[-] $K=1$ or vector operator with Euclidean components $\bs
A=(A_1,A_1,A_3)$ satisfying 
 \begin{equation} 
[J_j,A_k]=\rmi \hbar  \epsilon_{jkl} A_l, 
\end{equation} 
and spherical components $A_0=A_3$ and 
$A_{\pm1}=\mp\frac{1}{\sqrt{2}}(A_1 \pm \rmi A_2)$. 
\end{itemize}

\begin{cmexercise}{Irreducible or not?}

 Is the Hamiltonian $H=\frac{p^2}{2m} - \mu \bs
S\cdot \bs B$ of a free massive particle coupled to an
external magnetic field $\bs B$  via
the magnetic moment $\bs \mu = \mu \bs S$ of its spin $\bs S$ a scalar? 
An irreducible tensor operator? 
\end{cmexercise}

\subsubsection{State multipoles}
\label{statesmultipoles.sec}

In the usual Hilbert space basis, the statistical operator reads 
\eqlab{
\rho = \sum_{m,m'} \rho_{mm'} \ket{sm}\bra{sm'}. 
}{rhomm'} 
Here, the ket $\ket{sm}$ transforms under the irreducible
representation $\calD^{(s)}$, whereas the bra $\bra{sm'}$, as its
complex conjugate, transforms under $(\calD^{(s)})^*$, the
contragredient representation. The ket-bra $\ket{sm}\bra{sm'}$
transforms under the direct product $ \calD^{(s)}\otimes
(\calD^{(s)})^*$ which is reducible. 
One introduces therefore an ensemble of elements that do transform
under the irreducible representation $\calD^{(K)}$, 
\eqlab{
 \tkq := \tkq(s,s) := \sum_{m,m'} (-)^{s-m} \cg{ssm'\,{-m}}{KQ}  
\ket{sm'}\bra{sm}, 
}{deftkq}
with $K=0,1,\dots,2s$ and $Q=-K,-K+1,\dots,K$. 
The Clebsch-Gordan coefficients $\cg{s_1s_2m_1 m_2}{KQ}$ are the
coefficients of the unitary basis change 
from the direct product $\calH_{s_1} \otimes
\calH_{s_1}$ towards the Hilbert subspace $\calH_K$ 
of spin $K$ that should be familiar from the addition of two spins. 
The CG coefficients are
non-zero only if two selection rules are satisfied: 
$(i)$ the two magnetic quantum numbers on the left add up to the one
on the right, $m_1+m_2=Q$; $(ii)$ the angular momentum on the
right satisfies the triangle inquality $|s_1-s_2|\le K \le s_1+s_2$. 

In our case, we do not couple two spins, but a spin and its complex
conjugate. Since there is a contragredient representation in the game,
the CG coefficients in \refeq{deftkq} 
feature a characteristic minus sign in front of ${-m}$. 
The triangle selection rule implies for us $0\le K \le 2s$. 

The irreducible tensor operators 
$\tkq=:|KQ)$ form a basis of Liouville space that 
is properly orthonormal with
respect to the trace scalar product of matrices: 
\eqlab{
(KQ|K'Q') := \tr\{ (\tkq)^\dagger\tkqpr\} = \delta_{KK'}\delta_{QQ'} .
}{kqonb}
The Hermitian conjugate
is $(\tkq)^\dagger= (-)^Q T^{(K)}_{-Q}$. 
Any linear operator $A$ can be decomposed in this basis, 
\eqlab{
A = \sum_{KQ} A_{KQ} \tkq  \quad \text{with}\quad A_{KQ}:= (KQ|A) = \tr\{(\tkq)^\dagger
A\}.
}{decompA} 
The irreducible components 
\eq{
\rho_{KQ} = \tr\{(\tkq)^\dagger \rho \} = \expect{(\tkq)^\dagger}
}
of the density matrix are called 
\emph{state multipoles} or statistical
tensors, and have been introduced already in the 1950's by Fano and
Racah \cite{Blum,Fano}.  

\begin{cmexercise}{Irreducible tensor operators}

\noindent (0) Show that all $\tkq$ except
$T^0_0$ have zero trace and calculate the state monopole moment 
$\rho_{00}$ (use $\sqrt{2s+1}\cg{ssm'{-m}}{00} = (-)^{s-m}\delta_{mm'}$). 

\noindent (1) Show that the irreducible vector operator is proportional to the spin
operator, $T^1_Q=c_s^{-1/2}S_Q$. Hint: consider the action of $T^1_0$ 
($\sqrt{c_s}\cg{ssm \,{-m}}{10} = (-)^{s-m}m$), argue with rotational
invariance and fix the proportionality constant $c_s=s(s+1)d_s/3$ by computing
$\tr [(T^1)^2]$.  

\end{cmexercise}

\subsubsection{Irreducible spin superoperators}

By inserting the decomposition \refeq{decompA} on the left and right
side of an arbitrary superoperator acting like $\calL A$, one
obtains 
\eq{
\calL A = \sum_{KQK'Q'} |K'Q')\, (K'Q'|\calL|KQ) \,  (KQ|A)
}
such that with the notation $\calL_{K'Q',KQ}:=  (K'Q'|\calL|KQ)$ the
superoperator reads 
\eq{
\calL = \sum_{KQK'Q'}\calL_{K'Q',KQ} |K'Q') (KQ|   
}
where the Liouville-space dyadics on the right hand side transform under 
$\calD^{(K')}\otimes (\calD^{(K)})^*$, in an analogous manner to \refeq{rhomm'}. 
Following the same strategy as previously, we recouple the elements
again using the appropriate CG-coefficients to get irreducible
superoperators \cite{Mueller05}
\eq{
\calT^{L}_M (K,K'):=  \sum_{Q,Q'} (-)^{K-Q} 
\cg{K'KQ'{-Q}}{LM} \, |K'Q') (KQ| \ . 
}
Finally, any superoperator in completely decomposed form reads 
\eq{ 
\calL =\sum_{LM}\sum_{K,K'} \calL_{LM}(K,K') \calT^{L}_M (K,K')
} 
with coefficients 
$\calL_{LM}(K,K')=  \sum_{Q,Q'} (-)^{K-Q} 
\cg{K'KQ'{-Q}}{LM}\calL_{K'Q',KQ}$.
This decomposition is completely general and applies to arbitrary
superoperators. It is only worth the effort, however, if the
superoperator has rotational symmetries. The greatest gain in
computational speed and conceptual clarity is obtained if the
superoperator is a scalar such that  its only nonzero component is
$L=0, M=0$. In that case, which applies to our Lindbladian
\refeq{Lspinrelax}, one finds by virtue of the triangle rule that
$K=K'$: as promised, $\calL$ indeed connects subspaces of equal 
rank. We can choose the decomposition  
\eq{
\calL = \sum_{K=0}^{2s} \lambda_K \calT^{(K)}   
 } 
where the $\calT^{(K)}$ are orthogonal projectors onto the subspaces
$L(\calH_s)^{(K)}$ of irreducible tensor operators of rank $K$: 
\eq{
  \calT^{(K)} = \sqrt{2K+1}\calT^{0}_0 (K,K) = \sum_Q |KQ) (KQ| \ . 
}
They are indeed orthogonal, $\calT^{(K)}\calT^{(K')} = \delta_{KK'}
\calT^{(K)}$, 
by virtue of \refeq{kqonb} and resolve the identity in Liouville
space, 
$\sum_K \calT^{(K)} = 1$, by virtue of a completeness relation of
CG-coefficients. 

\subsubsection{Scalar relaxation rates}
\label{eigenvalues.sec} 

Once the invariant subspaces are known, the eigenvalues are obtained
by projecting the superoperator onto an arbitrary basis element: 
\eq{
\lambda_K = (KQ|\calbarL|KQ) = \tr \left\{(\tkq)^\dagger \calbarL\tkq \right\}.  
}
For the spin relaxation Lindbladian  \refeq{Lspinrelax}, one must
calculate commutators of the form $[S_i,[S_i,\tkq]]$. 
In order to use the defining commutation relations \refeq{ITOcom1} and
\refeq{ITOcom2} for irreducible tensor operators, one writes the
scalar product of spin operators in terms of the spherical components
$S_0=S_z$, 
$S_{\pm1}=\mp\frac{1}{\sqrt{2}}(S_x \pm \rmi S_y)$, 
\eq{
\sum_i S_i S_i = \sum_{p=0,\pm1} (S_p)^\dagger S_p = \sum_{p=0,\pm1}
(-)^p S_p S_{-p} 
}
such that 
\eq{
\lambda_K = -\frac{\gamma_s}{2s(s+1)} \sum_{p=0,\pm1} (-)^p
\tr\left \{(\tkq)^\dagger [S_{-p},[S_p,\tkq]]\right \}
} 
It is now a simple exercise to show with the help of 
\refeq{ITOcom1} and \refeq{ITOcom2} (by paying attention to the
supplementary factor $\sqrt{2}\hbar$ in the definition of the
raising/lowering components $ J_\pm = \sqrt{2}\hbar S_{\pm1}$) 
that the double commutator gives back the tensor operator itself, 
$ [S_{-p},[S_p,\tkq]] = c(p,Q,K) \tkq$. Summing all three terms gives
remarkably simple eigenvalues,  
\eqlab{
\lambda_K=-\gamma_s\frac{K(K+1)}{2s(s+1)}, \quad K=0,1,\dots,2s. 
}{lambdaK} 
These real, negative eigenvalues describe a pure relaxation process as
expected from the definition of the Lindbladian
\refeq{Lspinrelax}. 
They are scalar objects, also known as \emph{rotational invariants}, 
and can be
expressed in terms of $6j$-coefficients that are 
constructed out of the irreducible representations $\calD^{(K)}$ for
the state multipole, $\calD^{(s)}$ for
the spin $\bS$ itself and
$\calD^{(1)}$ for its coupling to the vector field $\bB$ 
\cite{BL81,Mueller05}. 
This type of consideration is of considerable importance in many different
fields of physics involving angular momentum or spin; for example,
relaxation coefficients very similar to \refeq{lambdaK}  characterise spatial correlations in certain
ground states of quantum spin chains 
\cite{Fannes89}.

\subsubsection{Isotropic spin relaxation}
\label{isotropic.spin.relax.sec}

The master equation $\partial_t\rho(t)= \calbarL \rho(t)$ separates
into uncoupled equations for each 
invariant subspace: 
\eq{
\partial_t\rho(t)= \sum_{KQ}\partial_t\rho_{KQ}(t)\tkq =
\sum_{KQ}\rho_{KQ}(t) \calbarL \tkq = \sum_{KQ}  
\rho_{KQ}(t)  \lambda_K \tkq
}
The resulting differential equation $\partial_t\rho_{KQ}(t)= \lambda_K
\rho_{KQ}(t)$ for the
state multipoles  is easily solved to yield a simple exponential decay  
\eqlab{
\rho_{KQ}(t) = \rme^{-\gamma_K t} \rho_{KQ}(0)
}{spinrelax.K} 
with relaxation rates $\gamma_K =|\lambda_K|$. 
This is a particular example for a state multipole relaxation as described by Blum 
in chapter 8 of \cite{Blum}.  
The
first two values deserve a special discussion. 
The scalar mode relaxation rate $\gamma_0=0$ assures the trace
preservation of $\rho(t)$. At the same time, this vanishing eigenvalue is
responsible for the fact that the Lindbladian is not invertible. 
The vector mode relaxation rate $\gamma_1=\gamma_s/[s(s+1)]=:1/\tau_1$ describes
the relaxation of the \emph{orientation} or average spin vector: 
\eq{
\expect{\bs S(t)}=\tr \{ \rho(t)\bS \} = \rme^{-t/\tau_1} \expect{\bs S(0)}
}  
since $\tr\{\tkq S_{Q'}\}$ projects onto $K=1$ (remember 
exercise 5.(1) and the orthogonality relation \refeq{kqonb}).

For a qubit spin $\frac{1}{2}$ this is all that needs to be calculated
since any $2\times 2$ density matrix can be parametrised as  
$\rho_2(t) = \frac{1}{2}\mathbbm{1}_2 + \expect{\bs S(t)} \cdot
\bs\sigma$ and thus 
\eq{
\rho_2(t) 
= \rme^{-t/\tau_1} \rho(0) + (1-\rme^{-t/\tau_1})
\frac{1}{2} \mathbbm{1}_2\ .
}
This isotropic spin  $\frac{1}{2}$ relaxation therefore is for all
times given by the depolarising channel \refeq{depolarchannel} with
$p_2(t)  = 1-\rme^{-t/\tau_1}$. The Kraus operators for the operator sum
representation \refeq{opsumrep} valid for \emph{all} times follow by
using $p_1(t) = \frac{3}{4}p_2(t)$: 
\eq{
W_0(t)    = \frac{\sqrt{1+3\rme^{-t/\tau_1}}}{2} \ \mathbbm{1}_2,
\qquad  
W_i(t)    = \frac{\sqrt{1-\rme^{-t/\tau_1}}}{2} \ \sigma_i,\quad i=1,2,3.
}
Naturally, 
by developing these expressions to first order in
$t/\tau_1=\frac{4}{3}\gamma_s t $ one finds 
the Kraus operators for an infinitesimal time step derived in
section \ref{quantumchannel.sec}. 
In general, it is easy to show by derivation that to each quantum dynamical
semigroup described by an exponential superoperator $\exp[\calL t]$
corresponds a Lindblad-type master equation \cite{Havel03},
cf.\ Sec.3.2.\ in \cite{klh06}.  
However, as always, the inverse
operation of
integrating the infinitesimal time evolution to finite times is much
harder.  
Deriving a set of Kraus operators for an arbitrary quantum channel in
general requires the complete diagonalisation of the microscopic
Hamiltonian.  
Luckily, for spin $\frac 1 2$ everything is so simple that the
complete calculation is possible. 

Naturally, one thus wonders whether the 
full-fledged angular momentum formalism is necessary at all to
describe isotropic spin relaxation. 
A simple calculation shows that in the ``depolarising channel'' defined
by \refeq{depolarchannel}, all non-scalar eigenvalues
$\lambda_1,\lambda_2,\dots$ are identical.
For that channel, one does not need to employ angular momentum
theory, and the Kraus operators are the generators of the
$\mathsf{su}(n)$ Lie algebra \cite{Ritter05}. 
But please be aware that this is \emph{not} the case for
our arbitrary spin coupled to a freely fluctuating impurity
spin where all higher state multipoles $K\ge 1$ come with their own
different decay rates \refeq{lambdaK}. 
For this channel, the author has not been able to determine the Kraus operators for
finite times (but would certainly be happy to receive any valuable
information on that point by his readers). 
In this case, there seems to be no easier way to describe the spin relaxation
than in terms of the irreducible components: 
\eqlab{ 
\rho(t)= \sum_{KQ} \rme^{-\gamma_Kt} \rho_{KQ}(0) \tkq. 
}{spinrelax.eq}
This implies the simple exponential decay  
\eqlab{
A_{KQ}(t) = e^{-\gamma_Kt}   A_{KQ}(0)
}{observable.relax} 
for the irreducible components \refeq{decompA} of any observable $A$.

\section{Diffusion}
\label{diffusion.sec}

\subsection{Transport} 
\label{transport.general} 

We may call ``transport'' a movement from a point $\br$ to a point
$\br'$ that is induced by an external cause. 
In free space, propagation is \emph{ballistic}: the average
square of the distance covered after a time $t$ scales like 
$\expect{r^2}\sim v^2 t^2$ where $v$ is the particle's velocity. 
A disordered medium contains \emph{impurities} that
interrupt the ballistic movement. So-called ``quenched'' disorder is fixed for each
realisation of an experiment, but varies from experiment to experiment
when samples are changed. Predictions about observables will
involve an average $\mv{(\dots)}$ by integrating over a classical
disorder distribution or by tracing out uncontrolled 
quantum degrees of freedom. Generically, the averaged expectation value
behaves as a diffusive
quantity: $\mv{\expect{r^2}}\sim 2 D t$ with $D$ the \emph{diffusion
  constant}.  

In a hydrodynamic description, 
diffusion is a direct consequence of two very basic hypotheses: 
\begin{enumerate}
\item[(i)] a local conservation law $\partial_t n +\bs\nabla \cdot \bs j
= 0$, also known as the
continuity equation, linking the local density $n(\br,t)$ and the
local current density $\bs j(\br,t)$, and 
\item[(ii)] a linear
response relation $\bs j= - D \bs\nabla n$, known as Fourier's law in the
context of heat transport and Fick's law in the context of particle diffusion.   
\end{enumerate}
By inserting the second relation into the first, one finds immediately
the diffusion equation
$(\partial_t -D\nabla^2)n = 0$. 
The hydrodynamic description is only valid for times and distances large
compared to the scales on which microscopic scattering takes
place. The linear response coefficient $D$ has to be
determined microscopically. 
In essence, the simplest physical process
leading to diffusion is a random walk or repeated elastic
scattering. A kinetic description where point particles collide with obstacles permits to
derive the diffusion constant $D$ associated with this process as
function of microscopic scattering parameters.  In this section, we will derive
the appropriate kinetic equation for elastic momentum scattering from first principles using a master
equation approach and determine the relevant 
diffusion constant. A largely equivalent presentation with more
details can be found in chapter 4 of the recommendable book ``Quantum
Transport Theory'' by J. Rammer \cite{Rammer}.  

\subsection{Momentum -- a primer on translations} 

If spin is to be understood by considering rotations, then momentum
is understood by considering translations. 

Let a particle be prepared at a point $\br_0=0$ with a spreading described by a
function $f(\br)$. After translation of the entire preparation apparatus by a
vector $\ba\in\R^d$, the new position is described by the
function   
\eqlab{
[T_\ba f](\br) := f(\br-\ba) 
}{translat.f}
such that the particle is indeed centered around $\br'_0=\ba$. In this
so-called ``active'' formulation of translations, relation \refeq{translat.f}
defines the action of the translation operator $T_\ba$ in a functional
space of, say, probability distributions, in exact analogy to the
case of rotations treated in section \ref{functrep.sec}.  
Here, it is an 
infinite-dimensional representation of the group $(\R,+)$ of real numbers with the addition ``+'' 
as a group law; to be precise, in $d$ dimensions it is the $d$-fold direct product of
such representations. Translations have the  
identity element $E=T_0$ and inverse
$T_\ba^{-1}=T_{-\ba}$. 
This group is Abelian because different translations commute: 
$ T_\ba T_\bb = T_{\ba+\bs b}= T_{\bb+\ba} = T_\bb T_\ba$. 
Furthermore, this group is a
simply connected Lie group, 
 and all translations can be generated by exponentiation 
$T_\ba = \exp\{-\rmi \ba\cdot\bT\}$ of $i=1,\dots, d$ generators $T_i$
that form a Lie algebra.%
\footnote{If, however, the possible positions lie on a lattice, only discrete
translations by lattice vectors are allowed, and translations are a
representation of $(\mathbbm{Z},+)$, the additive group of integer
numbers. In an infinite volume,
the translation group is not compact because it is not
bounded such that its representation theory
is quite different from the compact rotation groups SU$(n)$. Notably,
it has no finite-dimensional irreducible representations. 
An exception are  discrete translations
on a lattice with periodic boundary conditions, often used for
classification of crystalline lattices. This group is the cyclic
group, and its irreducible representations are labelled by the
admissible wave vectors of the reciprocal
lattice \cite{Cornwell97}. Physicists are taught to know this under the
epithet of Bloch's theorem.
 }

In the functional representation \refeq{translat.f}, all translations $T_\ba =
\exp\{-\rmi \ba\cdot\bT\}$ are generated by $\bT =-\rmi \bs\nabla$, the derivative with respect 
to the position coordinate, which becomes apparent through a Taylor series
expansion 
\eq{
f(\br-\ba) = f(\br) - \ba\cdot\bs \nabla f(\br) +\dots =
\exp\{-\ba\cdot\bs\nabla\} f(\br).
}
In quantum physics, 
the fundamental commutation relation between position
and momentum, $[\hat r_i,\hat p_j] = \rmi\hbar \delta_{ij}$, implies
that the momentum observable $\hat\bp= \hbar \bT$  is the translation
generator, $\hat\bp = -\rmi \hbar \bs\nabla$ in position
representation. 
Translations are implemented by
the unitary operator 
\eqlab{
U(\ba) = \exp\{ -\rmi \ba\cdot\hat\bp/\hbar\}
}{defUa}
such that the particle's position operator $\hat\br$ transforms as
\eq{
  \hat\br' = U(\ba)\hat\br U(\ba)^\dagger = \hat\br -\ba, 
} 
quite similarly to the corresponding identity \refeq{Uaction.rot} for
rotations. In contrast to the angular momentum commutation relations \refeq{commutationsu2},
the simpler version $[\hat p_i,\hat p_j]=0$ for the translation
generators reflects their Abelian structure.

The momentum operator is diagonal in the basis of momentum
eigenstates, $\hat\bp\ket{\bp} = \bp \ket{\bp}$. 
Any operator $O$ is translation invariant if and only if it 
commutes with the momentum operator, $[O,\hat\bp]=0$. Therefore,
it is diagonal in the momentum representation, $\bra{\bp'} O \ket{\bp} =
\delta_{\bp\bp'}O_{\bp}$. Notably, in
absence of any external perturbation, the Hamiltonian $H_0$ should
be translation invariant, $ H_0 \ket{\bp} =
\eps_\bp\ket{\bp}$, a property sometimes referred to by the expression
``the $p_i$'s are good quantum numbers''.  
Henceforth, we will choose units such that $\hbar=1$ and thus drop the
distinction between momentum and wave vectors: $\bp = \hbar \bk =
\bk$. 

The spatial form of the wave functions $\psi_\bp(\br)=
\bra{\br}\bp\rangle$ is fixed by their translational properties:
using \refeq{translat.f} together with \refeq{defUa} yields  
$\bra{\br-\ba}\bp\rangle = \bra{\br} [U(\ba)\ket{\bp}] = \exp\{-\rmi
\ba\cdot\bp\} \bra{\br}\bp\rangle$ such that $\bra{\br}\bp\rangle =
C \exp\{\rmi \br\cdot\bp\}$ up to a normalisation factor. In a finite
volume $\Omega=L^d$, these plane waves are square-integrable and can
be properly normalised. Instead, we choose to work in the limit $\Omega\to\infty$
and fix $C=1$. The identity is then 
resolved by 
\eq{
\mathbbm{1} = \int\rmd^d r \ket{\br}\bra{\br} =
\dint{p}\ket{\bp}\bra{\bp}. 
} 
Using the plane-wave expansion, Fourier transformation is written
using the convention   
\eq{
f(\bx) = \dint{p} \rme^{\rmi \bx\cdot\bp} f_\bp  \qquad \text{with}
\qquad 
f_\bp =  \int \rmd^d x \, \rme^{- \rmi \bx\cdot\bp} f(\bx). 
  }
This choice is convenient because factors of $2\pi$ are always
associated with $p$-integrals which, if required, can be
easily converted back to finite-volume sums, $\dint{p} F(\bp) ={L^{-d}}\sum_\bp F(\bp)$.

\subsection{Master equation approach to diffusion} 


We will show that a microscopic quantum derivation of diffusive behaviour is possible
starting from the single-particle Hamiltonian
\eq{
H = H_0 + V. 
}
Here the translation-invariant part $H_0$ describes free propagation in 
momentum eigenstates $\ket{\bp}$ with eigenenergies
$\eps_p=p^2/2m$; the generalisation to more general dispersion
relations is straightforward. 
The random impurity potential 
\eqlab{
V= \sum_{i} v(\hat\br - \bx_i) = \sum_{i}
\rme^{-\rmi\hat\bp\cdot\bx_i} v(\hat\br) \rme^{\rmi\hat\bp\cdot\bx_i} 
}{defVel} 
describes momentum scattering by an arbitrary potential $v(\br)$,
typically quite short-ranged, 
centered on random classical positions $\{\bx_i\}$. 
Deriving observable quantities will involve averages over
all possible realisations of the disorder. The ensemble average of any
quantity $O(\{\bx_i\})$ is the 
integral
\eqlab{
\mv{O} = \int(\prod_i \rmd^dx_i) P(\{\bx_i\}) O(\{\bx_i\})
}{ensemble-average}
over all inpurity positions weighted by their distribution $P(\{\bx_i\})$. 
The simplest distribution $P(\{\bx_i\})= \prod_i P(\bx_i)= \prod_{i=1}^N \Omega^{-1}$
describes $N$ 
uncorrelated impurities with average uniform density $n=N/\Omega$ in any finite
volume $\Omega$. This distribution will be used
in the following, and we will go to the
thermodynamic limit $N,\Omega\to\infty$ with fixed density $n$. 

This description is valid if the mass $M$ of scattering impurities is
much larger than the mass $m$ of scattered particles. This is the case
for electrons scattered by lattice
defects in solid state devices or for photons scattered by 
cold atoms when recoil can be neglected. The impurities then have no
internal dynamics and simply realise an external
potential $v(\br)$; this description can be  
obtained in the limit $m/M\to 0$ from the more general model,
where also the dynamics of impurities is taken into account. Note that
this limit is just the opposite of the usual picture used for
quantum Brownian motion (cf.\ Sec. 3.4 (c) in \cite{klh06}), where one tracks the movement of a large test
particle bombarded frequently by smaller ones. 

\subsubsection{Derivation of master equation} 
\label{deriv.meq.sec}
 
Here, we follow the standard derivation of a master equation for open
quantum systems \cite{GZ04,BP02} by adapting the Born-Markov  or
weak-coupling recipe to
our case, cf.\ Sec.\ 4.1 in \cite{klh06}. 
The ensemble average \refeq{ensemble-average} now plays the role  of a
trace over bath variables.  

Starting from the Liouville-von Neumann equation $\partial_t {\tilde \rho}(t) = -\rmi [\tilde
V(t),\tilde\rho(t)]$ in the interaction
representation $\tilde A(t) = U^\dagger_0(t) A U_0(t)$ and
developing to second order in the interaction leads to the 
pre-master equation for the averaged density matrix $\mv{\tilde\rho(\{\bx_i\},t)} =: \mv{\tilde\rho(t)}$:
\eqlab{
\partial_t \mv{\tilde{\rho}(t)} = -\rmi \mv{[\tilde V(t),\rho(0)]} - 
\int_0^t  \mv{\left[\tilde V(t),[\tilde V(t-t'),\tilde\rho (\{\bx_i\},t-t')]\right]}
\rmd t' 
. 
}{preME1}
We may assume that the initial density matrix $\rho(0)$ does not
depend on the disorder configuration -- it may for instance represent
an initial 
wave-packet prepared far from impurities whose temporal evolution we
wish to follow. Then, the first term $ [\mv{\tilde V(t)},\rho(0)]$ 
vanishes because 
\eqlab{
\mv{\tilde V(t)} = \tilde{\mv{V}}(t)= \mv{V} 
}{mvV.eq}
is a constant real number that shifts all energy levels $\eps_p$ of 
$H_0$ by a uniform amount and thus gives no contribution under the
commutator.  


The resulting equation is still exact, but not useful: it is
not a closed equation for the ensemble-averaged 
$\tilde{\bar{\rho}}(t)$ since the density matrix
inside the integral still depends on the disorder configuration. 
In order to cope with the integrand, one typically proceeds with the
so-called \emph{Born approximation}, replacing 
the exact density matrix inside the integral by its average:  
\eqlab{
\partial_t\tilde{\bar{\rho}}(t) =  - 
\int_0^t  \left[\mv{\tilde V(t),[\tilde V(t-t')},\tilde{\bar{\rho}} (t-t')]\right]
\rmd t'
\ . 
}{ME1}
Now, we are left with an effective Gaussian model of disorder since 
everything depends on the pair correlations $\mv{VV}$. 
Concerning the time dependence, we still face a difficult
integro-differential equation for $\tilde{\bar{\rho}}(t)$. If 
the time scale of scattering is much smaller than the time scale of
evolution we are interested in,
we can perform the \emph{Markov approximation} by replacing
$\tilde{\bar{\rho}}(t-t') \mapsto \tilde{\bar{\rho}}(t)$ inside the
integral and by letting the upper limit $t$ of integration go to $\infty$  
such that now we have a closed differential equation for 
$\tilde{\bar{\rho}}(t)$. Reverting to the Schr\"odinger representation
we find the following master equation for scattering 
by fixed impurities: 
\eqlab{
\partial_t{\bar{\rho}}(t) = - \rmi[H_0,\bar{\rho}(t)] + \calD\bar{\rho}(t) 
}{MEel} 
with a scattering superoperator $\calD$ defined by  
\eqlab{
\calD\bar{\rho}(t)  = \mv{V \bar{\rho}(t) W} + \mv{W \bar{\rho}(t) V} -
\mv{V W}\bar{\rho}(t) - \bar{\rho}(t) \mv{ WV}
}{calD}
where 
\eq{
W : = \int_0^\infty   \tilde V(-t')\rmd t' =  \int_{-\infty}^0 
U_0^\dagger(t') V U_0(t')\rmd t' =: \sum_j \rme^{-\rmi\hat\bp\cdot\bx_j} w(\hat\br)
\rme^{\rmi\hat\bp\cdot\bx_j}. 
} 
It will become apparent in section \ref{boltzmann.sec} that the weak
coupling or Born approximation discards genuine quantum
corrections and entails purely classical dynamics. Instead of
``approximation'', we had better speak of ``simplification'' because at this
stage we have no means of knowing whether the resulting description is
truly an approximation or perhaps qualitatively wrong. And really, 
in section \ref{wl.sec} we will see that quantum
corrections need to be considered in phase-coherent samples. 

\subsubsection{Momentum representation} 
\label{momentum.rep.sec}

In order to see what kind of evolution the master equation \refeq{MEel}  describes, we
evaluate it in the momentum representation in which the free
Hamiltonian $H_0$ is diagonal. In the short-hand notation
$\ket{1}\:=\ket{\bp_1}$ and 
$\eps_1=\eps_{\bp_1}$,  
the first scattering contribution reads
\eqlab{
\bra{1} \mv{V \bar{\rho}(t) W }\ket{4} 
= \sum_{2,3} \sum_{i,j} 
\mv{ \bra{1} \rme^{-\rmi\hat\bp\cdot\bx_i} v(\hat\br) \rme^{\rmi\hat\bp\cdot\bx_i}\ket{2} 
\bra{3} \rme^{-\rmi\hat\bp\cdot\bx_j} w(\hat\br) \rme^{\rmi\hat\bp\cdot\bx_j}\ket{4}
}   \bra{2}\bar{\rho}(t)\ket{3}. 
}{1VW4}
The terms with $i\neq j$ are
proportional to $\mv{V}^2$ and cancel with an equivalent
contribution in \refeq{calD} that comes with a minus sign (as before, the average
$\mv{V}$ gives no contribution thanks to the commutator structure of
the equation of motion). 
In the terms $i=j$, we can take the translation operators outside the matrix
elements and perform the ensemble average: 
\eqlab{
\begin{aligned}
\sum_i \mv{ \rme^{-\rmi (\bp_1 -\bp_2+\bp_3-\bp_4)\cdot\bx_i}} 
& = \frac{N}{L^d} \int \rmd^d x \, \rme^{-\rmi (\bp_1
-\bp_2+\bp_3-\bp_4)\cdot\bx} \\
& = n (2\pi)^d \delta(\bp_1+\bp_3-\bp_2-\bp_4) . 
\end{aligned}
}{momentumconservation}
As expected,  the average over a uniform distribution restores translational
invariance which is equivalent to the conservation of total momentum expressed
by $(2\pi)^d \delta(\bp_1+\bp_3-\bp_2-\bp_4) =: \delta_{1+3,2+4}$. 

Proceeding in the evaluation of \refeq{1VW4}, the first matrix element 
\eq{
\bra{1}v(\hat\br)\ket{2} =\int \rmd^d r \, \rme^{-\rmi
(\bp_1-\bp_2)\cdot\br} v(\br) =  v_{\bp_1-\bp_2}  
=: v_{12}
} 
is the Fourier transform of the scattering potential. 
In the second matrix element $\bra{3} w(\hat\br) \ket{4} =
\int_{-\infty}^0  \bra{3} U_0^\dagger(t') v(\hat\br)  U_0(t')  \ket{4} \rmd t'$, we can
pull out the time integration
\eqlab{
\int_{-\infty}^0 \rme^{\rmi(\eps_3-\eps_4)t'} \rmd t' = 
\frac{\rmi}{\eps_4-\eps_3+ \rmi 0 }
=:  \Gamma_{34}. 
}{defIeps} 
Readers with a background in perturbation theory 
will recognise this as the matrix element
$\Gamma_{34} = \rmi  \bra{\bp_3} G_0^\text{R}(\eps_4) \ket{\bp_3}$ of the
free retarded resolvent operator $G_0^\text{R}(\omega)=(\omega-H_0+\rmi 0)^{-1}$.   
This gives us a hint on the applicability of the Markov approximation:
the rapid time scale here is the inverse energy difference $\eps_3-\eps_4$
of incident and scattered state.  The effective evolution on much longer
time scales into a new state $\ket{\bp_3}$ is constrained by
\refeq{defIeps} to the energy shell $\eps_4$ of the incident state
$\ket{\bp_4}$.  
The imaginary contribution of $\Gamma$ produces the \emph{Lamb shift} that
renormalises the original energy levels, cf.\ Sec.\ 4.1 in \cite{klh06}, whereas the
real part yields the relaxation rates that render the dynamics 
irreversible.

Altogether, this first contribution to the collision functional reads 
\eqlab{
\bra{1} \mv{V\bar{\rho}(t) W }\ket{4} 
= n  \sum_{2,3}  \delta_{1+3,2+4} v_{12}v_{34}
\Gamma_{34} \bra{2}\bar{\rho}(t)\ket{3} . 
}{1VW4.final}
Collecting all four
contributions gives 
\eq{
\begin{aligned}
\bra{1} \calD \bar{\rho}(t)\ket{4} & = & n  \sum_{2,3} \Big[ 
\delta_{1+3,2+4} v_{12}v_{34}
(\Gamma_{12} + \Gamma_{34}) \bra{2}\bar{\rho}(t)\ket{3}  \\  
& &  - \delta_{2,3} (|v_{12}|^2 \Gamma_{21} + |v_{34}|^2 \Gamma_{43})
\bra{1}\bar{\rho}(t)\ket{4}  \Big ]. 
\end{aligned}
} 
We choose to use the parametrisation 
\eqlab{
\begin{aligned}
\bp_1 & = \bp+\bq/2,\quad     & \bp_2 & =  \bp'+\bq/2  \\ 
\bp_4 & = \bp-\bq/2,\quad    & \bp_3 & =  \bp'-\bq/2
\end{aligned}
}{kqparam.eq} 
that complies with the conservation of
total momentum \refeq{momentumconservation}. 
With this parametrisation, $v_{12}=v_{\bp-\bp'}=
v_{34}^*$. Furthermore, energy differences become 
\eqlab{
\eps_1-\eps_2 =: \eps_p-\eps_{p'} + \frac{(\bp-\bp')\cdot\bq}{2m}
}{eps12def} 
and $\eps_1-\eps_4 = \bp\cdot\bq/m$, and 
we write the sum of matrix elements \refeq{defIeps} in the form  
\eqlab{
\Gamma_{12} +\Gamma_{34} 
=: \Gamma_{\bp\bp'}(\bq) . 
}{Ikk'def}
Finally, disposing of the overbar $\bar{\rho}(t) \mapsto \rho(t)$, the 
density matrix elements $\rho(\bp,\bq,t) :=
\bra{\bp+\frac\bq 2}\rho(t)\ket{\bp-\frac \bq 2}$  obey the master equation 
\eqlab{
\partial_t \rho(\bp,\bq,t) = -\rmi \frac{\bq\cdot\bp}{m}
\rho(\bp,\bq,t) +  \calD[\rho(\bp,\bq,t)]
}{ME.kq}
with the scattering functional 
\eqlab{
 \calD[\rho(\bp,\bq,t)]= 
n \dint{p'} |v_{\bp-\bp'}|^2 
\left[ \Gamma_{\bp\bp'}(\bq) \rho(\bp',\bq,t) - \Gamma_{\bp'\bp}(\bq)
\rho(\bp,\bq,t) \right]. 
}{calDrhopq.def}

\subsubsection{Trace conservation  and continuity equation}

Clearly, the master equation  \refeq{ME.kq} has the form of a kinetic balance
equation where the scattering functional \refeq{calDrhopq.def} contains transitions 
$\bp'\to\bp$ that increase the magnitude of $\rho(\bp,\bq,t)$, and
also negative contributions of depleting transitions $\bp\to\bp'$. 
Since we have neither sinks nor external sources, the net effect must be zero which should be apparent in a
conservation of the local probability density. And really, the master
equation first of all preserves the trace, 
$\partial_t \tr\{\rho(t)\} = \dint{p}\partial_t\rho(\bp,0,t) = 0$,
since the collision functional is antisymmetric under the exchange
$\bp\leftrightarrow\bp'$ and thus  
\eq{ 
\dint{p}
 \calD[\rho(\bp,\bq,t)] =0. 
} 
This antisymmetry is inherited from the double-commutator
structure of \refeq{ME1} and holds for all spatial Fourier
momenta $\bq$.  
Summing the $\bq$-dependent master equation over
$\bp$ thus leads to the \emph{continuity equation} 
\eqlab{
\partial_t n_\bq(t) + \rmi \bq\cdot \bs j_\bq(t) = 0
}{continuity} 
that links the Fourier transforms of the
local density $n(\br,t)$ 
and local current density $\bs j(\br,t)$,  
given as the first two 
$\bp$-moments of the density distribution: 
\begin{subequations}
\begin{align} 
n_\bq(t)  & = \int\rmd^d r \, \rme^{-\rmi \bq\cdot\br}
n(\br,t)  =  \dint{p}  \rho(\bp,\bq,t), 
 \label{qdensity.eq} 
\\
\bs j_\bq(t) & = \int\rmd^d r \, \rme^{-\rmi \bq\cdot\br}
\bs j (\br,t)=   
\dint{p}  \frac{\bp}{m} \rho(\bp,\bq,t).  
\label{qcurrent.eq}
\end{align}
\end{subequations}
The current vanishes by parity for  isotropic distributions
$\rho(p,\bq,t)$. 

\subsubsection{Momentum isotropisation} 
\label{isotropisation.sec}  

What kind of dynamics does the master equation describe? A first,
simple answer is possible by considering the limit $\bq=0$ that
describes spatially averaged quantities. The definitions \refeq{defIeps},  \refeq{eps12def}
and \refeq{Ikk'def} imply $\Gamma_{\bp\bp'}(0) = 2\pi \delta(\eps_p-\eps_{p'})$
which assures the conservation of energy during elastic scattering. 
Since the isotropic energy $\eps_p$ fixes the modulus of
${\bp}'=p\hat\bn'$, the angular  
probability distribution $\rho(p\hat\bn,0,t) =: f_\eps(\hat\bn,t)$ at fixed energy  satisfies 
\eqlab{
\partial_t f_\eps(\hat\bn,t) = 2\pi n \dint{p'} \delta(\eps-\eps_{p'})
|v_{\bp-\bp'}|^2 \left[f_\eps(\hat\bn{}',t) - f_\eps(\hat\bn,t)\right]
. 
} {MEangle}
For an isotropic point-scatterer potential
$v(\br)=v_0 \delta(\br)$ that has no dependence on momentum, the
equation of motion takes the simple form 
\eqlab{
\partial_t f_\eps(\hat\bn,t) = -\gamel(\eps) \left[ 
f_\eps(\hat\bn,t) - \langle f_\eps(t)\rangle \right]   
}{ME.isotropic}  
where $\langle f_\eps(t)\rangle := \int\rmd \Omega'_d
f_\eps(\hat\bn{}',t)$ is 
the  angular average of the distribution, properly normalised: $\rmd \Omega'_d$
is $\rmd \theta'/2\pi$ in $d=2$ dimensions and $\rmd \phi' \rmd (\cos \theta')
/4\pi $ in $d=3$. 
The elastic scattering rate 
\eqlab{
\gamel(\eps) = 2\pi \nu(\eps) n v_0^2 
}{gammas} 
is defined in terms of the density of
states $\nu(\eps)= \dint{p'} \delta(\eps-\eps_{p'})$ and can equally
well be obtained  by Fermi's golden rule.   
Clearly, equation \refeq{MEangle} describes a simple exponential
decay of the initial angular distribution $f_\eps(\hat\bn,t=0)$  
towards a completely isotropic distribution  
$\langle f_\eps(t)\rangle $. 
Thus, our model of elastic momentum scattering by fixed
impurities leaves the kinetic energy conserved, but 
describes the isotropisation of the momentum distribution with a rate
$\gamel$.

By the same token, a
global net current $\bs j_0(t)$ 
 initially different from zero decreases to zero exponentially fast. 
 In other words, an initial wave packet launched with a
definite velocity loses the memory of its initial direction on a
time scale $\tauel=1/\gamel$.

\subsubsection{Boltzmann-Lorentz equation} 
\label{boltzmann.sec}

In order to know on what spatial scale the momentum isotropisation occurs and how the
average position of the wave packet evolves in time, we have to
consider the master equation \refeq{ME.kq} at finite $\bq$. We expect
diffusive behaviour to appear on large spatial scales and thus make a
Taylor expansion around $\bq=0$. Retaining only lowest-order terms in
$q/p \ll 1$ (which corresponds to spatial scales much larger than the
particle's wave length),   the equation of motion becomes 
\eqlab{
\partial_t \rho(\bp,\bq,t) + \rmi \frac{\bq\cdot\bp}{m}
\rho(\bp,\bq,t) = 
\calC[\rho(\bp,\bq,t)] 
}{ME.kq.elastic}
with the purely \emph{elastic}
collision integral 
\eqlab{
 \calC[\rho(\bp,\bq,t)]= 
2\pi n \dint{p'} \delta(\eps_p-\eps_{p'}) |v_{\bp-\bp'}|^2 
\left[\rho(\bp',\bq,t) - \rho(\bp,\bq,t) \right]. 
}{collrho.kq}
The only explicit occurrence of $\bq$ in \refeq{ME.kq.elastic} is now 
in the ballistic term on the
left-hand side that originates from the free
evolution with $H_0$. 

The parametrisation \refeq{kqparam.eq} has the additional advantage
that the density matrix elements 
%
$\rho(\bp,\bq,t)$ 
are the spatial Fourier
transform  of the Wigner distribution \cite{Hillery84} 
\eq{
\begin{aligned}
W(\bp,\br,t) & =  \frac{1}{(2\pi\hbar)^d} \int \rmd^d r'\bra{\br-\frac{\br'}{2}}
\rho(t) \ket{\br+\frac{\br'}{2}}\rme^{\rmi \bp\cdot \br'/\hbar} \\ 
& =\frac{1}{(2\pi\hbar)^d} \dint{q} \rme^{\rmi \bq\cdot\br}
\rho(\bp,\bq,t)
\end{aligned}
}
that represents the quantum density operator in a classical phase
space. Here, 
we  have momentarily restored $\hbar$'s visibility. With this standard
normalisation, the probability
density  $n(\br)=\bra{\br}\rho\ket{\br}$ with normalisation  $\int \rmd^dr n(\br) = 1$ is given as the marginal $n(\br) =
\int\rmd^d p W(\bp,\br)$. Conversely, the momentum distribution is 
$w(\bp) = \int \rmd^dr W(\bp,\br) = (2\pi\hbar)^{-d}
\rho(\bp,0)$, and $\int\rmd^dr\rmd^dp W(\bp,\br)=1$.

A Fourier transform with respect to $\bq$ now yields the
kinetic equation  
\eqlab{
\partial_t W(\bp,\br,t) +  \frac{1}{m} \bp\cdot\bs\nabla_{\br} W(\bp,\br,t)
=  \calC[W(\bp,\br,t)]
}{BL}
with the same elastic scattering integral \refeq{collrho.kq}.  
Remarkably, we have obtained precisely the linear Boltzmann-Lorentz equation for the
classical 
phase-space density $W(\bp,\br,t)$ under elastic scattering from fixed
impurities 
\cite{BalianII}. For photons, this equation is known as the radiative
transfer equation. 
Ex post, we can therefore conclude that the Born approximation 
\refeq{ME1} was the crucial step that discarded quantum corrections
to propagation amplitudes and left us with a classical phase-space
distribution. This  interpretation transpires also by analysing 
Feynman diagrams in perturbation theory and path-integral approaches 
(see chapter 4 of \cite{Rammer} for details). 

The \emph{weak disorder limit} in which the
Born approximation and Boltzmann transport theory are expected to be valid
corresponds to the regime where disorder corrections to the free
energy of the particle are small, $\gamel(\eps)\ll \eps$. 
The scattering time defines a typical length scale, the elastic
scattering mean free path $\elel= v_0 \tauel$ or average
distance between successive scattering events. The weak disorder limit can also be stated as 
$1/(k\elel)\ll 1$ which requires that successive scatterers are
placed in the scattering far field: the average distance
$\elel$ must be larger than the wavelength $\lambda=2\pi/k$. This is a low-density
argument because $\elel=1/(n\sigma_\text{el})$ in terms of the density
$n$ of  scatterers and their total elastic scattering cross section 
$\sigma_\text{el}$. 

It is allowed to neglect the explicit $\bq$-dependence inside
the collision integral \refeq{collrho.kq}, which contributes already a
factor $\gamel$, on hydrodynamic scales $q\elel\ll 1$. Consistently, it is
precisely in the hydrodynamic regime that we wish to determine the
diffusion constant. We have already derived the continuity equation
\refeq{continuity}. Making step (ii) of the general argument
presented in \ref{transport.general}, we now turn to
the calculation  
of the diffusion coefficient in the linear
response regime.

\subsection{Linear respose and diffusion constant} 

Any phase-space distribution $W_\text{eq}(p)$ that is homogeneous, independent
of time and rotation invariant, i.e., depends only on the modulus of $\bp$, is 
a solution of the Boltzmann-Lorentz equation \refeq{BL} since each
term vanishes separately. 
The corresponding density 
matrix elements are of the form $\rho_\text{eq}(\bp,\bq,t)=(2\pi)^d\delta(\bq)\rho_\text{eq}(p)$. 
This type of
solution is called a \emph{global equilibrium}. The underlying
statistics could be a 
Fermi-Dirac or Bose distribution, or their classical limit, the
Boltzmann distribution. 

By creating a small gradient of concentration, one can then 
induce a linear-response current that is  proportional
to the driving gradient; the coefficient of proportionality is the
diffusion constant. 
Kinetic theory permits to calculate linear response
coefficients. We will follow the linearisation method developed by
Chapman and Enskog \cite{BalianII} in order to derive the diffusion
constant, but in terms of the density matrix components $\rho(\bp,\bq,t)$ instead of
the space-dependent Wigner distribution because it will
prove useful to work with  Fourier-transformed quantities.

\subsubsection{Linearisation \`a la Chapman-Enskog} 
\label{ChapEns.sec} 

Suppose that initially, the distribution function $\rho(p,\bq,0)$ is
a \emph{local equilibrium} solution (i.e., 
isotropic in momentum) established by local scattering on a
rapid time scale $\tauel=1/\gamel$. However, we assume a non-delta like 
dependence on $\bq$, i.e., a finite gradient in real
space. It is therefore no longer a global 
equilibrium solution of the Boltzmann-Lorentz equation. The time
$\tau_\text{eq}$ it takes to reach global equilibrium is much longer than the local
scattering time such that we have a small parameter $\tauel/\tau_\text{eq} \ll 1$
(sometimes called the ``Knudsen number'').  
The linearisation method of Chapman and Enskog works by 
expanding  the distribution function formally in powers of this small parameter, 
\eq{
\rho(\bp,\bq,t) = \rho_0(\bp,\bq,t) + \tauel \rho_1 (\bp,\br,t) +
O(\tauel^2). 
} 
For the linear response calculation, these first two terms will suffice. 
The collision integral effectively multiplies the distribution by $\gamel= 1/\tauel$
such that 
$\calC[\tauel^n \rho_n]=O(\tauel^{n-1})$. Identifying equal
orders on both sides of the kinetic equation, we find to order $\tauel^{-1}$: 
\eq{
0 = \calC[\rho_0]
} 
which is satisfied if $\rho_0(p,\bq,t)$ is
\emph{locally} isotropic. By parity, the current density \refeq{qcurrent}  
then is entirely generated by the correction, 
\eqlab{
\bs j_\bq(t) = \dint{p} \frac{\bp}{ m}\tauel(\eps_p) \rho_1(\bp,\br,t). 
 }{jlinresp} 

The continuity equation \refeq{continuity}  then implies that to
lowest order 
the local density remains time-independent:  $\partial _t
\rho_0(p,\br,t)=0$. To this order $\tauel^0=1$, the
kinetic equation reduces to  
\eqlab{
\rmi\frac{\bq\cdot \bp}{m} \rho_0(p,\bq)
=  \calC[\tauel \rho_1(\bp,\bq)]. 
}{kineq.order.g1}
In order to calculate the current \refeq{jlinresp}, we need to solve
this equation for $\rho_1$ as function of $\rho_0$. We
content ourselves with the simple case of an isotropic point-scattering
potential $v_\bp=v_0$ treated in section \ref{isotropisation.sec}
 (other cases can be treated by an expansion in
angular eigenfunctions, see \cite{LeBellac} for the case of a
potential with scattering anisotropy). Then, using the right-hand
side of \refeq{ME.isotropic}, we find 
\eq{
\calC[\tauel\rho_1] = - \rho_1(\bp,\bq)
}
plus an isotropic term $\expect{\rho_1}$ that does not contribute to the
current \refeq{jlinresp} anyway and can be dropped such that 
\eqlab{
\rho_1(\bp,\bq) = - \rmi \frac{\bq\cdot\bp}{m}\rho_0(p,\bq). 
}{distribution}

\subsubsection{Diffusion coefficient} 

Inserting \refeq{distribution} into 
\refeq{jlinresp}, the resulting current reads 
\eq{
\bs j_\bq = -\frac{\rmi}{m^2} \dint{p}  \bp (\bq\cdot\bp) \tauel(\eps_p)
\rho_0(p,\bq).  
}  
By isotropy, $\int\rmd^d p  \, p_i p_j 
f(p) = (\delta_{ij}/d) \int\rmd^d p \,  p^2 f(p) $, and 
the current is collinear with $\bq$. To lowest order in $q$ 
we find
\eqlab{
\bs j_\bq = -\rmi \bq D_0 n_\bq,  
}{currentLR} 
the Fourier-transformed version of the linear response relation
$\bs j = - D_0 \bs\nabla n$. 
Kinetic theory has allowed us to to calculate the diffusion coefficient  
\eq{
D_0 = \frac{\mv{v}_0^2 \mv{\tau}}{d}
} 
as the product of an effective velocity and scattering time
averaged over the momentum distribution, 
\eq{ 
\mv{v}_0^2 \mv{\tau} = \frac{1}{m^2} \dint{p} p^2 \tauel(\eps_p) \rho_0(p)
.  }
Often, the distribution $\rho_0(p)$ is a sharply peaked function
around a certain momentum $p_0$ (for instance the Fermi momentum
$p_\text{F}$ for electrons), whereas $p^2 \tauel(\eps_p)$ according to \refeq{gammas} 
varies smoothly with
the density of states such that $\mv{v}_0^2 \mv{\tau} = p_0^2 \tauel(\eps_{p_0})/m^2$.  
In terms of the scattering mean-free path $\elel$, the
diffusion constant for isotropic point scatteres can also be
written 
\eqlab{ 
D_0 = \frac{\mv{v}_0\elel}{d}. 
}{D0} 
For anisotropic scattering, one has to replace the scattering
mean-free path $\elel$ by the transport mean-free path $\eltr$
\cite{LeBellac}. 

\subsubsection{Diffusion} 

Inserting the linear response  current  \refeq{currentLR} in
the continuity equation leaves us with a simple differential equation
$\partial_t n_\bq(t)= -D_0 q^2 n_\bq(t)$ 
that is immediately integrated to give an 
exponential decay 
\eqlab{
n_\bq(t) = \rme^{-D_0q^2t} n_\bq(0)
}{diffdens} 
for the Fourier components of the initial density fluctuation. Long-range fluctuations
($q\to 0$) take a very long characteristic time $\tau_q=1/(D_0 q^2)\to\infty$ 
to relax because of the constraint imposed by local
conservation. This diagonal decomposition into
Fourier modes with their continuous momentum index $\bq$ is the
analog of the spin relaxation 
\refeq{observable.relax}  where the discrete index $K$ separates high-$K$
irreducible modes with rather large relaxation rates
$\gamma_K$ from the isotropic component $K=0$ or trace that is
conserved.
\footnote{This analogy becomes even clearer if positions in a finite
volume are restricted to a lattice of $L^d$ sites, because the
irreducible representations of the discrete translation group (which under 
periodic boundary conditions  is the $d$-fold direct product of the
cyclic group $\mathbbm{Z}_L$)  
are precisely labelled by the different allowed $\bq$-vectors of the
reciprocal lattice \cite{Cornwell97}. }

The expectation value of the average radius squared, 
\eq{
\mv{\expect{r^2}}(t) =
\tr\{r^2\bar{\rho}(t)\}= -\nabla_q^2 n_\bq(t)|_{\bq=0},
}
for the diffuse density \refeq{diffdens} reads: 
\eqlab{
\mv{\expect{r^2}}(t) = \expect{r^2}_0 + 2 d D_0 t. 
}{diffradius} 
As expected, the long-time behavior of the particle's displacement
is indeed governed by the Boltzmann diffusion constant \refeq{D0}.

\section{Diffusive spin transport}

Having described in section \ref{spinrelaxation.sec} the relaxation of
a single motionless spin in a fluctuating field, and 
in section \ref{diffusion.sec} the diffusion of spin-less
massive particles, we now combine these two pictures and consider
a spin on a massive carrier particle that moves and encounters
impurities. The Hamiltonian is still of the form 
$H = H_0 + V$
where $H_0$ describes ballistic propagation in momentum and spin eigenstates
$\ket{\bp \sigma }:=\ket{\bp}\otimes \ket{s \sigma }$ with
spin-independent eigenenergies 
$\eps_{\bp} = p^2/2m$ (spin quantum numbers will from now be called $\sigma$ in
order to avoid confusion with the particle's mass). 
The impurity potential $V$ could describe momentum scattering, spin-flip
scattering, spin-orbit coupling, and the like, by randomly distributed
scatterers. 

In the following, we will consider in detail the case of 
freely orientable magnetic impurities that induce \emph{spin-flips}. Other mechanisms
such as spin-orbit scattering can be treated along the 
same lines. Actual
laboratory realisations include electronic spin-flip scattering, quite
relevant even for very low impurity concentrations, and the
randomisation of photon polarisation under the influence of scattering
by  atoms with degenerate dipole
transitions in cold atomic clouds.

\subsection{Master equation approach to diffusive spin transport}

\subsubsection{Deriving the master equation}

In addition to the elastic momentum scattering potential
\refeq{defVel} that for clarity we now call $V_\text{el}$, consider
then a spin-flip interaction potential  
\eqlab{
\Vsf= \sum_{j} v_\text{sf}(\hat\br-\bx_j) \bS \cdot \bs\tau_j
}{defVsf}
between the spin $\bS$
and a collection of freely orientable magnetic
impurities modelled as spin $\frac{1}{2}$  with Pauli matrices $\bs\tau_j$ centered at sites $\bx_j$. The
magnitude of spin-spin coupling has been  included in the
 short-ranged potential $v_\text{sf}(\br)$ whose spatial dependence induces momentum
scattering. The ensemble average now contains the usual average
\refeq{ensemble-average} over
random sites as well as a trace $\tr_{\{\tau\}}(\rho_{\{\tau\}}
\cdot)$ 
over impurity spins. We will assume that these impurities are 
distributed independently and isotropically, $\rho_{\{\tau\}}= \bigotimes_j
\rho_j$ with $\rho_j= \frac{1}{2}\mathbbm{1}_2$. 

It is now a simple task to derive a spin-diffusion master equation
for the density matrix $\rho(t)$ that operates on the combined Hilbert space
$\calH= \calH_s\otimes \calH_p$ of spin and momentum
by retracing exactly the steps of sections \ref{deriv.meq.sec} and
\ref{momentum.rep.sec} with the new potential $V= V_\text{el} +
V_\text{sf}$ instead of just $V_\text{el}$. 
Within the Born approximation second order in $V$, 
mixed terms $\mv{V_\text{sf}V_\text{el}}= \mv{V_\text{sf}}\;\mv{V_\text{el}}$
give no contribution (just as the product of averages before), and we can
consider the impact of $V_\text{sf}$ separately. A typical term arising in the new 
spin-flip part is  the 
counterpart of \refeq{1VW4}, $\bra{1}\mv{V_\text{sf} \rho(t)
W_\text{sf}}\ket{4}$. 
Together with the momentum-scattering factors appearing
already in \refeq{1VW4.final}, we find now an additional sum over
spin indices that defines the action of a spin-flip
superoperator 
\eqlab{ 
\begin{aligned}
\bra{\sigma_1}\calV \rho\ket{\sigma_4} & = \sum_{\sigma_2,\sigma_3}
\mv{ \bra{\sigma_1} \bS\cdot\bs\tau\ket{\sigma_2}  \bra{\sigma_3}
\bS\cdot\bs\tau\ket{\sigma_4}} \bra{\sigma_2} \rho\ket{\sigma_3}\\
& =  \sum_{\sigma_2,\sigma_3}
\bS_{\sigma_1\sigma_2}\cdot\bS_{\sigma_3\sigma_4} \bra{\sigma_2} \rho\ket{\sigma_3}
 \end{aligned}
}{defcalVsf} 
where the isotropic average over the impurity spin leads to the scalar
contraction $\frac{1}{2} \tr_\tau\{ \tau_i\tau_j\} =\delta_{ij}$
(cf.~exercise 1). 

Collecting all terms then gives the master equation that describes
elastic and spin-flip scattering in the small-$\bq$ limit: 
\eqlab{
\partial_t \rho(\bp,\bq,t) +\rmi \frac{\bq\cdot\bp}{m} \rho(\bp,\bq,t)= \calC[\rho]
+\calL[\rho] 
}{MEsf} 
Here, the elastic collision integral 
\eqlab{
\calC[\rho] = \frac{\gamma}{\nu(\eps)} \dint{p'}
\delta(\eps-\eps_{p'}) \left[\rho(\bp',\bq,t) -
\rho(\bp,\bq,t)\right]
}{calCtot}
describes momentum isotropisation  with a total scattering rate 
\eqlab{
\gamma = \gamel+\gamma_\text{sf}}
{ratesum}
that includes the spin-flip contribution
$\gamma_\text{sf} =  s(s+1) 2\pi
\nu(\eps)v_\text{sf}^2 n_\text{sf}$.   
Summing the rates, a prescription known as ``Mathiessen's rule''
\cite{Ashcroft-Mermin}, is permitted if the scattering mechanisms do not
interfere with each other which is the case in the low-density Born
approximation considered here. Note that the collision integral
\refeq{calCtot}
acts solely on the momentum degrees of freedom, but is the identity
in spin space. Genuine spin-flips  are generated by  
\eq{
\calL[\rho] =   \frac{1}{\nu(\eps)} \dint{p'}
\delta(\eps-\eps_{p'}) \calbarL\rho(\bp',\bq,t)
} 
where $\calbarL=\gamma_\text{sf}\hat\calL$ is precisely the spin relaxation Lindbladian 
\refeq{Lspinrelax} derived in
section \ref{lindbladian.sec}, now with the spin-flip relaxation rate
$\gamma_\text{sf}$.  

\subsubsection{Diffusive spin relaxation}
\label{spindiffusion.sec} 

Solving the spin-flip master equation \refeq{MEsf} in the linear
response regime is now exceedingly simple by projecting it onto irreducible spin
components $\rho^{(K)}=\sum_Q\rho_{KQ}\tkq$ since the spin relaxation
Lindbladian is diagonal in that basis, $ \calbarL \rho^{(K)} = \lambda_K
\rho^{(K)}$, with eigenvalues $\lambda_K$ given by \refeq{lambdaK}. 

Summing the master
equation for the $K$th spin sector over $\bp$  gives the continuity 
equation 
\eq{
\partial_t n_\bq^{(K)} (t) +\rmi \bq\cdot\bs j_\bq^{(K)}(t) = -\gamma_K
n_\bq^{(K)} (t)
}
for the density $n_\bq^{(K)}=\dint{p}\rho^{(K)}(\bp,\bq)$ and
associated current density $\bs j_\bq^{(K)}$. The spin-flip
Lindbladian is responsible for the appearance of a source term on the
right hand side, or rather a sink with spin relaxation rate
\eq{
\gamma_K = |\lambda_K| = \gamma_\text{sf} \frac{K(K+1)}{2s(s+1)}. 
}   
In the limit $\bq\to 0$, we recover exactly the global spin relaxation
\refeq{spinrelax.K}
of section \ref{isotropic.spin.relax.sec}. In particular, the total
trace is conserved since the spin trace is the
$K=0$ sector with vanishing eigenvalue $\lambda_0=0$.  

In the linear response regime, the Chapman-Enskog method of section
\ref{ChapEns.sec} carries through in each spin sector $K$. A difference
occurs for the time derivative of the locally isotropic component 
$\rho_0^{(K)}(p,\bq,t)$. The master equation \refeq{MEsf} implies  
\eq{
\partial_t \rho_0^{(K)}(p,\bq,t) = -\gamma_K  \rho_0^{(K)}(p,\bq,t)
} 
which is solved by $\rho_0^{(K)}(p,\bq,t) = \rme^{-\gamma_Kt} \rho_0^{(K)}(p,\bq,0)$
instead of just being constant as in \ref{ChapEns.sec}. 
Finally, the spin-sector density components show diffusive as well
as spin-relaxation dynamics 
\eq{
n_\bq^{(K)}(t) = \rme^{-D_0 q^2 t} \rme^{-\gamma_Kt} n_\bq^{(K)}(0)
}
where the Boltzmann diffusion constant $D_0$ is evaluated with the total momentum
relaxation time $\tau=1/\gamma$ from \refeq{ratesum}. 
  
Now we can answer the question that was the starting point of the
lecture (cf.\ Fig.\ \ref{spindiff.fig}): 
Imagine that we can inject spin-polarised particles 
$\ket{\up}:=\ket{s,\sigma =+s}$ with probability 
$p_\up(0)=\bra{\up}\rho_0\ket{\up}=1$  
on one end of a diffusive medium of length $L$. What is the
probability $p_\up(L)$ of retaining the spin polarisation at the other
end, assuming that we have spin-sensitive detection?   

By taking the matrix elements $\bra{\up}\rho^{(K)}(t)\ket{\up}$ 
of each irreducible spin component, we find that  
 the probability  relaxes during the transmission time $t = L^2/2D_0$
as 
\eq{
\begin{aligned}
p_\up(t) & = \frac{1}{2s+1} +\frac{3s^2}{s(s+1)(2s+1)} \rme^{-t/\tau_1}+
  \dots \\
 & =   \frac{1}{2}(1+\rme^{-t/\tau_1})
\end{aligned}
}
with the last line valid for electrons for which $1/\tau_1=
4\gamma_\text{sf}/3$. 
Equivalently, the \emph{degree of spin polarisation} 
\eq{
\pi(t) = \frac{p_\up(t)-p_\down(t)}{p_\up(t)+p_\down(t)} =
\rme^{-t/\tau_1}
}
simply relaxes from unity to zero. 
Naturally, for long enough times, the
distibution relaxes towards its 
equilibrium value $p_\text{eq} =\frac{1}{2s+1}$ to have any 
magnetic quantum number $ \sigma= -s,\dots,s$.   

In terms of length scales, the spin relaxation time permits to define
the spin relaxation length $\lambda_\text{sf} = \sqrt{2D_0
\tau_\text{sf}}$. In solid-state devices, the density of magnetic
impurities can be controlled such that $\gamma_\text{sf}\ll\gamel$
which means that spin coherence can be maintained quite efficiently,
even on scales $L\gg \elel$ where the momentum dynamics is no longer ballistic, but
already diffusive.

\subsection{Quantum corrections} 

In the Boltzmann transport theory
developed in section \ref{diffusion.sec}, one propagates effectively 
classical probabilities. However, in an environment that preserves the
phase coherence of the wave, one must propagate probability
\emph{amplitudes} that, by 
the superposition principle, allow for
interference phenomena. 
Elastic impurity scattering does preserve the phase coherence of the
propagating wave which means that we have to expect quantum
corrections to the Boltzmann transport theory whenever external
phase-breaking mechanism are so rare that the corresponding dephasing
time $\tauphi$ is much longer than the elastic scattering time
$\tauel$. 

\subsubsection{Weak localisation} 
\label{wl.sec}

\begin{figure}
\begin{center}
\includegraphics[width=0.8\textwidth]{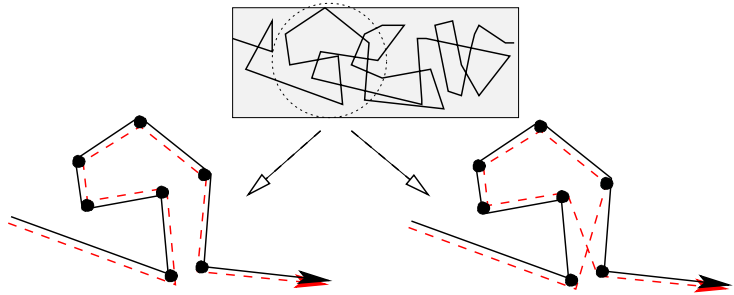}
\caption{\itshape \small Diffusive path (above) with the classical Boltzmann contribution of
co-propagating amplitudes (left) and the corresponding
counter-propagating amplitudes (right) that lead to weak
localisation corrections. 
}
\label{wl.fig}
\end{center}   
\end{figure}

A prominent example for such a quantum correction is weak localisation
(WL): the resistance of weakly disordered metallic
samples shows  a negative magnetoresistance $\partial \rho/\partial
B>0$ at small fields if spin-orbit scattering is absent
\cite{Bergmann84,Zwerger98}. 
This contradicts the classical Boltzmann-Drude 
picture that predicts that the resistance should increase with a magnetic field. 
The reason can be understood by considering the quantum return
probability to a point which includes the constructive interference of
counter-propagating amplitudes and is therefore larger than the
classical probability, see figure \ref{wl.fig}. The interference effect is masked by a
large enough magnetic field since loops of different sizes pick up
different Aharonov-Bohm phases.   

Without any dephasing mechanism nor an external magnetic field, 
the interference correction $D= D_0 + \Delta D$ to the 
classical diffusion constant $D_0$ 
is given by summing the contributions of all closed diffusive paths,  
\eq{
\frac{\Delta D}{D_0} = -\frac{1}{ \pi \hbar \nu(\epsilon) }
\dint{q} \int_0^\infty \rmd t \rme^{- D_0 q^2 t } \propto \dint{q}
\frac{1}{ D_0 q^2 }. 
}
In low dimensions $d=1,2$, this $q$-integral diverges for small $q$
which indicates that WL corrections become very important whenever
the interfering amplitudes can explore large length scales. 
The formal divergence is cured by a cutoff that can be 
either the system size itself or a finite phase-coherence length
$L_\phi = \sqrt{D_0\tau_\phi}$ 
due to dephasing mechanisms on a time scale $\tau_\phi$. 
\footnote{The ultraviolet divergence for large $q$ is cut off by the
shortest scale of scattering, $\elel$, and irrelevant for the present considerations.} 

\subsubsection{Dephasing of weak localisation by spin-flip scattering}

Including spin-flip scattering into the weak-localisation picture can be done by
diagonalising the spin-flip vertex appropriately \cite{Mueller05}. The irreducible
subspaces turn out to be the usual singlet and triplet state 
subspaces spanned by the Hilbert-space vectors $\ket{KQ}$
\footnote{Not operators $|KQ)$ in Liouville space as for the propagated intensity: the
difference is due to the fact that the two amplitudes in weak
localisation loops propagate in opposite directions such that two
states have to be recoupled instead of a state and its complex
conjugate.}. 
To sketch the result, the WL correction is written 
\eq{
\frac{\Delta D}{D_0} = -\frac{1}{ \pi \hbar \nu(\epsilon) }
 \dint{q} \sum_{K=0}^{2s} \frac{w_K}{D_0 q^2 +  \tau_c(K)^{-1}} 
}    
where each spin channel comes with a weight 
$w_K = (-)^{2s+K}(2K+1)/(2s+1)$. 
For electrons, $w_0=-\frac{1}{2}$ and $w_1=\frac{3}{2}$. 
More importantly, each spin channel is damped with its coherence time
$\tau_c(K)$.  They are given in terms of the spin relaxation rates
\refeq{lambdaK} by a recoupling relation that reduces to 
\eqlab{
\frac{1}{\tau_c(K)} = \frac{2}{\tausf} + \lambda_K  =
\frac{2}{\tausf}\left(1-\frac{K(K+1)}{4s(s+1)}\right)
}{tckoftausf} 
such that for electrons $\tau_c(0)=\tausf/2$ and
$\tau_c(1)=3\tausf/2$.   

\begin{figure}
\begin{center}
\includegraphics[width=0.65\textwidth]{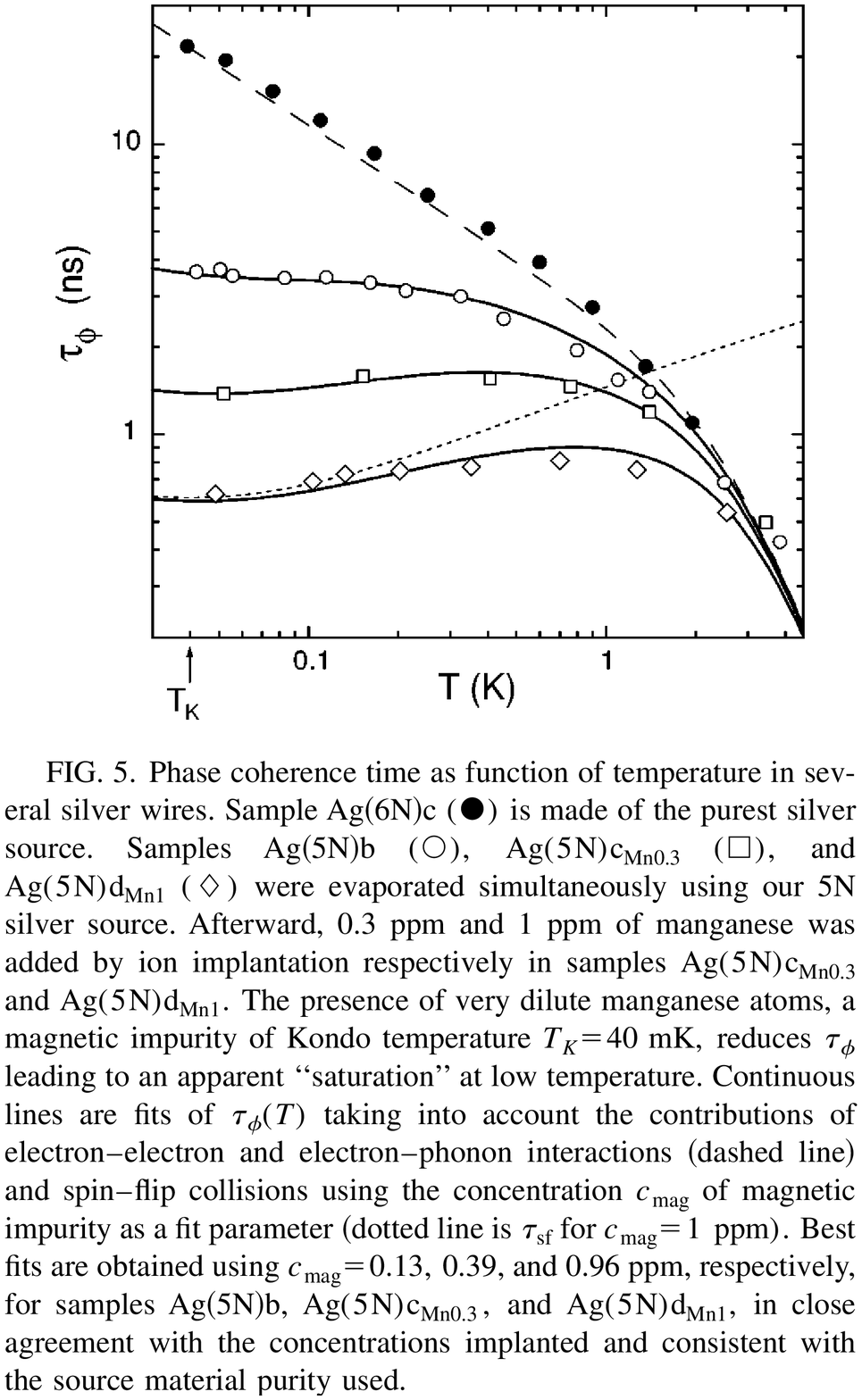}
\caption{\itshape \small 
This ``set of experiments suggests that the frequently
observed `saturation' of $\tau_\phi$ in weakly 
disordered metallic thin films can be attributed to spin-flip
scattering from extremely dilute magnetic impurities, 
at a level undetectable by other means'';  reprinted with permission from F. Pierre et
al., Phys. Rev. B \textbf{68}, 085413 (2003). Copyright (2003) by the
American Physical Society. 
}
\label{fig5pierre.fig}

\end{center}   
\end{figure}

Clearly, the interference in all spin channels is dephased rather
efficiently with a rate given essentially by the spin-flip rate, the
numerical prefactor being entirely fixed by geometry. 
This effect is visible even in quite pure samples as
admirably shown by F. Pierre and coworkers \cite{Pierre03}, see figure
\ref{fig5pierre.fig}.
 The complete theory needed for linking the experimental data to impurity concentrations
requires to take into account the interaction between
electronic excitations near the magnetic impurity (Kondo physics) as
well as the temperature-dependent dynamics of impurity spins 
(Korringa physics) that have both been neglected within the present
lecture; for details, see
\cite{Pierre03}.

Multiple coherent
scattering of photons in clouds of laser-cooled atoms is also subject
to spin-flip physics: the so-called coherent backscattering effect (an
interference enhancement of backscattered intensity) is severely
reduced when photons are scattered by dipole transitions with a Zeeman
degeneracy \cite{Labeyrie03c}. The analytical theory for multiple
coherent scattering of polarised photons by degenerate dipole
transitions employs the concepts of irreducible
decompositions exposed in section
\ref{spinrelaxation.sec}. Compared to electrons, the theory appears much
simpler because at low light intensity, photons do not interact. 
However, the treatment of photon propagation is more 
involved because the field transversality adds another source of polarisation relaxation
that needs to be taken into account \cite{Mueller02}.

Thus, magnetic impurities  are a very efficient source of dephasing
for interference of spin-carrying particles: the large ground-state
degeneracy implied by the random orientations of freely orientable
impurity spins permits dephasing even at zero temperature -- when other
decoherence processes like electron--phonon or electron--electron
scattering are suppressed --
because no
energy exchange is involved and stocking which-path information comes
for free. In return, whenever the impurity degrees of freedom can be
constrained by other means, then perfect
coherence is restored. This has been shown in Aharonov-Bohm
interference experiments with electronic samples subject to a strong external magnetic field that 
aligns the impurity spins \cite{Washburn86,Pierre02}. 
Similarly, in atomic clouds an external magnetic fields
that lifts the internal atomic Zeeman
degeneracy can be used to enhance the effective phase coherence
length of diffusing photons \cite{Sigwarth04}. 

\clearpage 

\subsection*{Acknowledgements}

It is a pleasure to thank the organisers and participants of the
summer school for providing a stimulating atmosphere of scientific
exchange in which numerous questions and remarks from students and colleagues 
permitted me to refine my arguments. Sincere thanks to Klaus Hornberger for accompanying my first
steps into Kraus representations, to R.F.~Werner for bringing
some of the mathematical background and reference \cite{Fannes89} to
my attention, to Michael Wolf for further clarifying discussions, and
to Hugues Pothier for helpful remarks concerning weak localisation
measurements in mesoscopic electronic devices. Finally, I am grateful
to the \'Ecole normale sup\'erieure (Paris) for its hospitality that helped me
to finalise these lecture notes and to Christian Wickles for a careful
reading of the manuscript.



\begin{thebibliography}{99}

\bibitem{klh06} 
K. Hornberger, ``Introduction to decoherence theory'', Lect. Notes Phys. \textbf{768}, 221-276 (2009); 
arXiv:quant-ph/0612118 

\bibitem{Zutic04}
I. $\check{\mathrm Z}$uti\'c, J. Fabian, and S. Das Sarma, 
``Spintronics: Fundamentals and applications'', 
Rev. Mod. Phys. \textbf{76}, 323 (2004)

\bibitem{CH}
M. Chaichian and R. Hagedorn, \textit{Symmetries in Quantum Mechanics}
(IOP Publishing, Bristol, 1998)

\bibitem{BL81} 
L.C. Biedenharn and J.D. Louck, ``Angular Momentum in Quantum
Mechanics'', vol. 8 of the \textit{Encyclopedia of Mathematics}
(Addison Wesley, 1981)

\bibitem{UG25}
G.E. Uhlenbeck and S. Goudsmit, ``Ersetzung der Hypothese vom
unmechanischen Zwang durch eine Forderung bez\"uglich des inneren
Verhaltens jedes einzelnen Elektrons'', Naturwiss. \textbf{13}, 953 (1925) 

\bibitem{Pauli25} 
W. Pauli, ``\"Uber den Zusammenhang des Abschlusses der
Elektronengruppen im Atom mit der Komplexstruktur der Spektren'', Z. Phys. \textbf{31}, 765 (1925) 

\bibitem{GS22}
W. Gerlach and O. Stern, ``Der experimentelle Nachweis der
Richtungsquantelung im Magnetfeld'', Z. Phys. \textbf{9}, 349 (1922) 


\bibitem{Gross95} D.J. Gross, ``Symmetry in Physics: Wigner's legacy'',
Physics Today (December 1995) p. 46, and
  references therein. 

\bibitem{Cartan13} 
E. Cartan, ``Les groupes projectifs qui ne laissent invariante aucune
multiplicit\'e plane'', Bull. Soc. Math. France \textbf{41}, 53 (1913) 

\bibitem{Weyl50} H. Weyl, \textit{The Theory of Groups and Quantum
Mechanics} (Dover Publications, 1950), first published as 
\textit{Gruppentheorie und Quantenmechanik} in 1928.  

\bibitem{Zeh}
H.-D. Zeh, \textit{The Physical Basis of the Direction of Time},
chap. 3 
(Springer) 

\bibitem{vKampen}
N. van Kampen, \textit{Stochastic Processes in Physics and Chemistry}
(North Holland, Amsterdam, 1990)

\bibitem{NielsenChuang} 
M.A. Nielsen and I.L. Chuang, \textit{Quantum Computation and Quantum
Information} (Cambridge University Press, 2002) 

\bibitem{Ritter05} 
W.G. Ritter, ``Quantum Channels and Representation Theory'', arXiv: quant-ph/0502153


\bibitem{FS} 
E. Fick and G. Sauermann, \textit{The Quantum Statistics of Dynamic
  Processes} (Spinger Series in Solid-State Science vol. 86, 1990)

\bibitem{Zwanzig64}
R. Zwanzig, ``On the identity of three generalized master equations'',
Physica \textbf{30}, 1109 (1964)

\bibitem{Gabriel}
H. Gabriel, 
``Theory of the influence of environment on the angular distribution
of nuclear radiation'', 
Phys. Rev. \textbf{181}, 506  (1969), especially the appendix. 

 
\bibitem{Preskill} J. Preskill, 
\href{http://www.theory.caltech.edu/people/preskill/ph229/}
{\textit{Lecture Notes for Physics 229:}}
\textit{Quantum Information and Computing}, chap. 3 (1998),\\
\url{http://www.theory.caltech.edu/people/preskill/ph229/}

\bibitem{GZ04} 
C.W. Gardiner and P. Zoller, \textit{Quantum Noise} (Springer, 2004) 


\bibitem{Mukamel04}
S. Mukamel, 
``Superoperator representation of nonlinear response: Unifying
quantum field and mode coupling theories'', 
Phys. Rev. E \textbf{68}, 021111 (2004)


\bibitem{Blum} 
K. Blum, \textit{Density Matrix Theory and Applications} 
(Plenum Press, 1996)

\bibitem{Fano} 
U. Fano and G. Racah, \textit{Irreducible Tensorial Sets} (Academic,
New York); U. Fano, ``Description of States in Quantum Mechanics by
Density Matrix and Operator Techniques'', Rev. Mod. Phys. \textbf{29}, 74 (1957).


\bibitem{Mueller05} 
C.A. M\"uller, C. Miniatura, E. Akkermans, and G. Montambaux,
``Mesoscopic scattering of spin $s$ particles'', 
J. Phys. A: Math. Gen. \textbf{38}, 7807 (2005). 

\bibitem{Fannes89} 
M. Fannes, B. Nachtergaele, and R.F. Werner, 
``Exact antiferromagnetic ground states of quantum spin chains'', 
Europhys. Lett. \textbf{10}, 633-637 (1989)  

\bibitem{Havel03} 
T.F. Havel, ``Robust procedures for converting among Lindblad, Kraus
and matrix representations of quantum dynamical semigroups'',
J. Math. Phys. \textbf{44}, 534 (2003) 


\bibitem{Rammer}
J. Rammer, \textit{Quantum Transport Theory}  (Perseus Books, Reading, 1998)


\bibitem{BP02} 
H.-P. Breuer and F. Petruccione, \textit{The Theory of Open Quantum
Systems} (Oxford 2002)


\bibitem{Cornwell97} 
J.F. Cornwell, \textit{Group Theory in Physics. An Introduction}
(Academic Press, 1997)

\bibitem{Hillery84} 
M. Hillery, R.F. O'Connell, M.O. Scully, and E.P. Wigner, ``Distribution
function in physics: fundamentals'', 
Phys. Rep. \textbf{106}, 121-167 (1984) 

\bibitem{BalianII} 
R. Balian, \textit{From Microphysics
to Macrophysics} vol. II (Springer, 1992)  

\bibitem{LeBellac} 
M. Le Bellac, F. Mortessagne, and G. Batrouni, \textit{Equilibrium and Nonequilibrium
Statistical Thermodynamics} (Oxford, 2002)  

\bibitem{Ashcroft-Mermin} 
N.W. Ashcroft and N.D. Mermin, \textit{Solid State Physics} (Harcourt
Brace, 1976)


\bibitem{Bergmann84} 
G. Bergmann, ``Weak localization in this films: a time-of-flight
experiment with conduction electrons'', Phys. Rep. \textbf{107}, 1
(1984)

\bibitem{Zwerger98} 
W. Zwerger, ``Theory of Coherent Transport'', 
in: T. Dittrich et al., \textit{Quantum Transport and Dissipation} (Wiley-VCH, 1998)

\bibitem{Pierre03} 
F. Pierre, A.B. Gougam, A. Anthore, H. Pothier, D. Est\`eve, and 
N. Birge, ``Dephasing of electrons in mesoscopic metal wires'', 
Phys. Rev. B \textbf{68}, 085413 (2003) 

\bibitem{Labeyrie03c}   G. Labeyrie, D. Delande, C.A. M\"uller,
C. Miniatura, and R. Kaiser, 
``Coherent backscattering of light by cold atoms: Theory meets experiment'', 
Europhys. Lett. \textbf{61}, 327 (2003)

\bibitem{Mueller02} {C. A. M{\"u}ller and C. Miniatura}, 
``Multiple scattering of light by atoms with internal degeneracy'', 
J. Phys. A: Math. Gen. \textbf{35}, 10163 (2002)  	

\bibitem{Washburn86} 
S. Washburn and R. Webb, ``Aharonov-Bohm effect in normal metal: 
Quantum coherence and transport'',  Adv. Phys. \textbf{35}, 375 (1986) 

\bibitem{Pierre02} 
F. Pierre and N. Birge, 
``Dephasing by Extremely Dilute Magnetic Impurities Revealed by
Aharonov-Bohm Oscillations'', 
Phys. Rev. Lett. \textbf{89}, 206804 (2002)

\bibitem{Sigwarth04} 
O. Sigwarth,  G. Labeyrie, T. Jonckheere, D. Delande, R. Kaiser, and 
C. Miniatura, ``Magnetic field enhanced coherence length in cold
atomic gases'', 
Phys. Rev. Lett. \textbf{93}, 143906 (2004)

\end{thebibliography}
\end{document}